\documentclass[11pt]{article}
\pdfoutput=1
\usepackage{jheppub}
\usepackage{subcaption}
\DeclareMathOperator{\Tr}{Tr}

\DeclareMathOperator{\Det}{Det}

\newcommand{\ri}{{\rm i}}

\def\cob{\delta}

\newcommand{\hf}{\frac{1}{2}}
\newcommand{\qu}{\frac{1}{4}}

\def\til#1{\widetilde{#1}}
\def\si{\sigma}

\def\del{\partial}

\def\bra{\langle}
\def\ket{\rangle}

\def\la{\lambda}

\def\ka{\kappa}
\def\h#1{\widehat{#1}}

\def\ga{\gamma}
\def\Ga{\Gamma}

\def\om{\omega}

\def\rt#1{\sqrt{#1}}

\newdimen\tableauside\tableauside=1.0ex
\newdimen\tableaurule\tableaurule=0.4pt
\newdimen\tableaustep
\def\phantomhrule#1{\hbox{\vbox to0pt{\hrule height\tableaurule width#1\vss}}}
\def\phantomvrule#1{\vbox{\hbox to0pt{\vrule width\tableaurule height#1\hss}}}
\def\sqr{\vbox{%
  \phantomhrule\tableaustep
  \hbox{\phantomvrule\tableaustep\kern\tableaustep\phantomvrule\tableaustep}%
  \hbox{\vbox{\phantomhrule\tableauside}\kern-\tableaurule}}}
\def\squares#1{\hbox{\count0=#1\noindent\loop\sqr
  \advance\count0 by-1 \ifnum\count0>0\repeat}}
\def\tableau#1{\vcenter{\offinterlineskip
  \tableaustep=\tableauside\advance\tableaustep by-\tableaurule
  \kern\normallineskip\hbox
    {\kern\normallineskip\vbox
      {\gettableau#1 0 }%
     \kern\normallineskip\kern\tableaurule}%
  \kern\normallineskip\kern\tableaurule}}
\def\gettableau#1{\ifnum#1=0\let\next=\null\else
\squares{#1}\let\next=\gettableau\fi\next}

\title{Wilson loops in 3d $\mathcal{N}=4$ SQCD from Fermi gas}

\author{Kazumi Okuyama}

\affiliation{Department of Physics, 
Shinshu University, Matsumoto 390-8621, Japan}

\emailAdd{kazumi@azusa.shinshu-u.ac.jp}

\abstract{
We study 1/2 BPS Wilson loops in 3d $\mathcal{N}=4$ $U(N)$
Yang-Mills theory with one adjoint and $N_f$ fundamental hypermultiplets
from the Fermi gas approach.
By numerical fitting, we find the first few worldsheet instanton corrections
to the Wilson loops with winding numbers 1, 2 and 3.
We verify that our Fermi gas results are consistent with
the  matrix model results in the planar limit.
}

\begin{document}

\maketitle

\renewcommand{\thefootnote}{\arabic{footnote}}
\setcounter{footnote}{0}
\setcounter{section}{0}

\section{Introduction}

Fermi gas approach \cite{MP}
is a powerful technique to
study large $N$ behavior of 
the partition functions of
various three dimensional theories.
Of particular interest is the so called ABJ(M) theory
\cite{Aharony:2008ug,Aharony:2008gk}
which describes the worldvolume theory on $N$ M2-branes
on the orbifold $\mathbb{C}^4/\mathbb{Z}_k$.
It turns out that the partition function 
of ABJ(M) theory on a three sphere
correctly reproduces
the $N^{3/2}$ scaling of free energy
expected from the holographically dual M-theory
on $AdS_4\times S^7/\mathbb{Z}_k$ \cite{Drukker:2010nc}.
The Fermi gas formalism enables us to go further 
beyond this perturbative behavior, and now we have
a complete understanding of the instanton corrections
to the free energy coming from wrapped M2-branes on the bulk M-theory side
(see \cite{Hatsuda:2015gca,Marino:2016new} for reviews).
Also, Fermi gas formalism can be applied to the computation of Wilson loops
in ABJ(M) theory and many interesting properties were uncovered
in the last few years
\cite{KMSS,HHMO,Matsumoto:2013nya,Hatsuda:2016rmv,Matsuno:2016jjp,Okuyama:2016deu,Kiyoshige:2016lno}.
However, for many theories with less supersymmetry than the ABJ(M) case,
we are still lacking a detailed understanding of the large $N$ behavior of
free energy.  

In \cite{Mezei:2013gqa,Grassi:2014vwa,HO},
some progress has been made in the so called $N_f$
matrix model which describes
the $S^3$ partition function of $d=3$ $\mathcal{N}=4$
$U(N)$ Yang-Mills theory coupled to one adjoint and $N_f$
fundamental hypermultiplets.
This theory naturally appears as the worldvolume
theory of $N$ D2-branes in the presence of $N_f$ D6-branes,
which in turn is holographically dual to
M-theory on $AdS_4\times S^7/\mathbb{Z}_{N_f}$
in the large $N$ limit. Here the $\mathbb{Z}_{N_f}$
action on $S^7$ is induced from the orbifold 
$\mathbb{C}^2\times (\mathbb{C}^2/\mathbb{Z}_{N_f})$
with $A_{N_f-1}$ ALE singularity.
By the 3d $\mathcal{N}=4$ mirror symmetry,
this theory is dual to a $\h{A}_{N_f-1}$ quiver gauge theory \cite{Intriligator:1996ex,deBoer:1996mp}.
As emphasized in \cite{Grassi:2014vwa},
the $N_f$ matrix model admits not only the ordinary  't Hooft limit
(type IIA limit on the bulk side)
\begin{align}
 N,N_f\to\infty\quad\text{with}~~\la=\frac{N}{N_f}:~\text{fixed},
\label{eq:thooft-limit}
\end{align}
but also the M-theory limit
\begin{align}
 N\to\infty\quad\text{with}~~N_f:~\text{fixed}.
\label{eq:M-limit}
\end{align}
Here $1/N_f$ plays the role of string coupling.
An important consequence of the Fermi gas
description is that these two limits are actually smoothly connected
since the grand partition function of $N_f$ matrix model is a completely well defined
quantity
for any value of $N_f$.
In \cite{HO}, the first few instanton corrections to the grand potential
were determined in a closed form as a function of $N_f$;
it is found that the structure of instanton corrections of 
$N_f$ matrix model is quite different from the 
ABJM case, and in particular there is no obvious connection 
with the topological string.
However, the pole cancellation mechanism between worldsheet instantons and membrane instantons,
originally found in the ABJM case \cite{Hatsuda:2012dt}, works also in the $N_f$ matrix model \cite{HO}.

In this paper, we will consider the Wilson loops in the $N_f$ matrix model from the Fermi gas
approach. We will focus on the 1/2 BPS Wilson loop on $S^3$ which wraps $m$ times around the equator.
By the numerical fitting, we find the first few worldsheet instanton corrections
to the vacuum expectation value (VEV) of winding Wilson loops with winding number
$m=1,2,3$.
We find that our Fermi gas result is consistent with the
planar VEV of winding Wilson loops in the 't Hooft limit \eqref{eq:thooft-limit}
obtained from the resolvent of $N_f$ matrix model \cite{Grassi:2014vwa}.
Our Fermi gas result suggests that there is no ``pure'' membrane instanton corrections
to the Wilson loop VEVs except for the contributions
from bound states of membrane instantons and worldsheet instantons.
This is reminiscent of the 
instanton corrections to the 1/2 BPS Wilson loops in ABJ(M)
theory \cite{HHMO,Hatsuda:2016rmv}.

This paper is organized as follows.
In section \ref{sec:W-gas}, we consider VEVs of 1/2 BPS winding
Wilson loops in $N_f$
matrix model from the Fermi gas approach
and explain our numerical method
to compute them.
In section \ref{sec:pert}, we find the perturbative part of winding
Wilson loops in a closed from, and in section \ref{sec:inst} we determine the first few worldsheet
instanton corrections for the Wilson loops 
with winding number $m=1,2,3$.
In section \ref{sec:wkb}, we reproduce the perturbative part of
Wilson loop in the fundamental representation from 
the WKB expansion
of spectral traces.
In section \ref{sec:planar}, we summarize the planar solution of
$N_f$ matrix model obtained in \cite{Grassi:2014vwa}.
We find that the planar resolvent in \cite{Grassi:2014vwa}
can be vastly simplified.
In section \ref{sec:tHooft},
we compare the Fermi gas results and the matrix model results of the
Wilson loop VEVs in the planar limit and find a perfect agreement.
Finally, we conclude in section \ref{sec:conclude}.
In appendix \ref{app:exact} we list some exact values of Wilson loop VEVs 
and in appendix \ref{app:curious}
we mention some curious properties of Wilson loops for $N_f=4$.

\section{Winding Wilson loops}\label{sec:W-gas}

\subsection{Review of $N_f$ matrix model}
First we review the $S^3$ partition function of $d=3$ $\mathcal{N}=4$ 
$U(N)$ Yang-Mills theory with one adjoint and $N_f$ fundamental
hypermultiplets.
In $d=3$, the gauge coupling has mass dimension $1/2$ and
it flows to infinity in the IR.
Thus the gauge kinetic term is irrelevant in IR
and the $S^3$ partition function
of this theory is independent of the gauge coupling.
This theory flows to a superconformal fixed point in the IR, which  is 
conjectured to be holographically dual to 
M-theory on $AdS_4\times S^7/\mathbb{Z}_{N_f}$.

Using the supersymmetric localization, 
the $S^3$ partition function is reduced to a finite dimensional integral \cite{Kapustin:2009kz}, 
which is dubbed the $N_f$ matrix model in \cite{Grassi:2014vwa}
\begin{align}
\begin{aligned}
 Z(N,N_f)&=\frac{1}{N!}\int\frac{d^Nx}{(4\pi)^N}\prod_{i=1}^N\frac{1}{(2\cosh\frac{x_i}{2})^{N_f}}\prod_{i<j}\tanh^2\frac{x_i-x_j}{2}.
\end{aligned}
\label{eq:Zint}
\end{align}
Using the Cauchy identity, \eqref{eq:Zint}
can be rewritten as
a partition function of $N$ fermions on a real line
\begin{align}
 Z(N,N_f)
=\frac{1}{N!}\int d^N x\sum_{\si\in S_N}(-1)^\si\prod_{i=1}^N\rho(x_i,x_{\si(i)})
\end{align}
where the density matrix $\rho$ is given by
\begin{align}
 \rho(x,y)=\frac{1}{2\pi}\frac{1}{(2\cosh\frac{x}{2})^{N_f/2}}\frac{1}{2\cosh\frac{x-y}{2}}\frac{1}{(2\cosh\frac{y}{2})^{N_f/2}}.
\end{align}
It is also useful to express
$\rho(x,y)$ as a matrix element of 
the quantum mechanical operator
$\h{\rho}$
\begin{align}
\begin{aligned}
 \rho(x,y)=\bra x|\h{\rho}|y\ket,\qquad 
 \h{\rho}=\frac{1}{(2\cosh\frac{\h{x}}{2})^{N_f/2}}\frac{1}{2\cosh\frac{\h{p}}{2}}
\frac{1}{(2\cosh\frac{\h{x}}{2})^{N_f/2}},
\end{aligned}
\label{eq:rho-QM}
\end{align}
with $\h{x}$ and $\h{p}$ being the 
coordinate and momentum of a fermion
obeying the canonical commutation relation
\begin{align}
 [\h{x},\h{p}]=\ri\hbar,\qquad\hbar=2\pi.
\label{eq:hbar-2pi}
\end{align}

As discussed in \cite{MP},
it is more convenient to consider 
the grand canonical ensemble
by summing over $N$ with fugacity $\ka=e^\mu$. 
It turns out that the grand partition function
is written as a Fredholm determinant
\begin{align}
 \Xi(\mu,N_f)=1+\sum_{N=1}^\infty \ka^N Z(N,N_f)
=\Det(1+\ka\rho)
\end{align}
which in turn is physically interpreted as a grand canonical ensemble of ideal Fermi gas \cite{MP}.
The large $N$ behavior of canonical partition function
$Z(N,N_f)$ can be deduced from
the large $\mu$ behavior of the 
\textit{modified grand potential} $J(\mu,N_f)$, which is related to the grand partition function by
\cite{Hatsuda:2012dt}
\begin{align}
 \Xi(\mu,N_f)=\sum_{n\in\mathbb{Z}}e^{J(\mu+2\pi\ri n,N_f)}.
\end{align}
Then the canonical partition function is recovered from
 $J(\mu,N_f)$ by the integral transformation
\begin{align}
 Z(N,N_f)=\int_{\mathcal{C}}\frac{d\mu}{2\pi\ri} e^{J(\mu,N_f)-N\mu}
\label{eq:mu-int}
\end{align}
where $\mathcal{C}$ is a contour on the $\mu$-plane
running from $e^{\frac{\pi\ri}{3}}\infty$ to $e^{-\frac{\pi\ri}{3}}\infty$.

The modified grand potential $J(\mu,N_f)$ can be decomposed into several pieces:
\begin{align}
 J(\mu,N_f)= J^{\text{pert}}(\mu,N_f)+J^{\text{WS}}(\mu,N_f)+J^{\text{M2}}(\mu,N_f)+J^{\text{bound}}(\mu,N_f)
\label{eq:decompJ}
\end{align}
The first term of \eqref{eq:decompJ}
is the perturbative part
\begin{align}
 J^{\text{pert}}(\mu,N_f)=\frac{C\mu^3}{3}+B\mu+A,
\label{eq:Jpert}
\end{align}
where $C,B,$ and $A$ are given by \cite{Mezei:2013gqa,Grassi:2014vwa,HO}
\begin{align}
 \begin{aligned}
  C=\frac{2}{N_f\pi^2},\qquad
B=\frac{1}{2N_f}-\frac{N_f}{8},\qquad
A=\hf A_c(N_f)+\hf A_c(1)N_f^2.
 \end{aligned}
\label{eq:abc}
\end{align}
Here $A_c(N_f)$ is a certain
resummation of the constant map contributions of topological string
\cite{Hanada:2012si,Hatsuda:2015owa,HO}
\begin{align}
 A_c(N_f)=-\frac{N_f^2\zeta(3)}{8\pi^2}
+4\int_0^\infty dx\frac{x}{e^x-1}\log\Bigl(2\sinh\frac{2\pi x}{N_f}\Bigr).
\end{align}
As shown in \cite{HO},
$A_c(N_f)$ can be evaluated in a closed form
for integer $N_f$. In particular, for $N_f=1$ we find
\begin{align}
 A_c(1)=-\frac{\zeta(3)}{8\pi^2}+\qu\log2.
\end{align}
$J^{\text{WS}}(\mu,N_f)$ and $J^{\text{M2}}(\mu,N_f)$
in \eqref{eq:decompJ} denote the worldsheet instanton and membrane
instanton corrections, respectively,
while the last term
$J^{\text{bound}}(\mu,N_f)$ in \eqref{eq:decompJ}
represents the contributions from
the bound states of worldsheet instantons and membrane instantons. 
In \cite{HO}, the first few instanton corrections are determined: 
the worldsheet instantons are given by
\begin{align}
 \begin{aligned}
  J^{\text{WS}}(\mu,N_f)&=-\frac{4\mu+N_f}{2\pi\sin\frac{2\pi}{N_f}}e^{-\frac{4\mu}{N_f}}\\
&+\left[-\frac{(4\mu+N_f)^2}{4\pi^2}
+\frac{3\sin\frac{6\pi}{N_f}(8\mu+N_f)}{8\pi\sin\frac{2\pi}{N_f}\sin\frac{4\pi}{N_f}}
-\frac{\sin\frac{8\pi}{N_f}}{2\sin^2\frac{2\pi}{N_f}\sin\frac{4\pi}{N_f}}\right]e^{-\frac{8\mu}{N_f}}
+\mathcal{O}(e^{-\frac{12\mu}{N_f}})
 \end{aligned}
\end{align}
and the membrane 1-instanton is given by
\begin{align}
 J^{\text{M2}}(\mu,N_f)=-\frac{\Ga\bigl(-\frac{N_f}{2}\bigr)^2}{4\pi^2\Ga(-N_f)}(2\mu+1)e^{-2\mu}
+\mathcal{O}(e^{-4\mu}).
\end{align}
At present, we do not know the exact form of the bound states
in the $N_f$ matrix model.

In the large $N$ limit, the integral \eqref{eq:mu-int}
can be evaluated by the saddle point approximation, where
the saddle point value $\mu_*$  of the chemical potential 
is given by
\begin{align}
 \mu_*\approx \rt{\frac{N}{C}}=\pi\rt{\frac{NN_f}{2}}.
\label{eq:mu-saddle}
\end{align}
Plugging this value $\mu_*$ into the perturbative part of grand potential,
we recover the $N^{3/2}$ behavior of free energy
\begin{align}
 -\log Z(N,N_f)\approx N\mu_*-J^{\text{pert}}(\mu_*,N_f)\approx\frac{\pi\rt{2N_f}}{3}N^{3/2}.
\label{eq:N3/2}
\end{align}
Also, the free energy receives instanton corrections of order
\begin{align}
 \text{worldsheet~1-instanton}&:~e^{-\frac{4\mu_*}{N_f}}=e^{-2\pi\rt{2N/N_f}},\\
\text{membrane~1-instanton}&:~e^{-2\mu_*}=e^{-2\pi\rt{N_fN/2}}.
\label{eq:order-inst}
\end{align}

\subsection{Wilson loops in $N_f$ matrix model}
In this paper, we will study the VEV of 1/2 BPS Wilson loops in
the $N_f$ matrix model.
The 1/2 BPS Wilson loop in representation $R$ is given by
\cite{Assel:2015oxa}\footnote{1/2 BPS Wilson loops in $\mathcal{N}=4$
Chern-Simons-matter theories are studied in \cite{Cooke:2015ila}.}
\begin{align}
 \Tr_R\mathcal{P}\exp\left[\oint ds 
\bigl(A_\mu \dot{x}^\mu +\si|\dot{x}|\bigr)\right]
\end{align}
where $\si$ is one of the three scalar fields
in the $\mathcal{N}=4$ vectormultiplet 
and $x^\mu(s)$ parametrizes the equator of $S^3$.
The VEV
of such BPS Wilson loops
can be reduced to a finite dimensional integral 
by the supersymmetric localization \cite{Kapustin:2009kz}.
Here we will focus on the Wilson loop wound around the equator
of $S^3$ $m$ times, which we will call the winding Wilson loop with winding number $m$.

Now, let us consider the (un-normalized) VEV of winding Wilson loop\footnote{
We define the Wilson loop VEV without dividing by the dimension $N$
of representation
\begin{align}
W_m\not=\frac{1}{N}\Biggl\bra \sum_{i=1}^N e^{mx_i}\Biggr\ket
\end{align}
Our definition will simplify the grand canonical VEV of $W_m$.
}\label{foot:normalization}
\begin{align}
 W_m(N,N_f)=\Biggl\bra
\sum_{i=1}^Ne^{mx_i}\Biggr\ket
\end{align}
where the expectation value is defined by
\begin{align}
 \bra\mathcal{O}\ket=\frac{1}{N!}\int\frac{d^Nx}{(4\pi)^N}\,
\mathcal{O}\prod_{i=1}^N\frac{1}{(2\cosh\frac{x_i}{2})^{N_f}}\prod_{i<j}\tanh^2\frac{x_i-x_j}{2}.
\label{eq:matrix-vev}
\end{align}
Note that the integral defining $W_m(N,N_f)$ is convergent 
if $N_f$ satisfies the condition
\begin{align}
 N_f>2m.
\label{eq:Nf-cond}
\end{align}
%It is also useful to defined the normalizaed VEV
%\begin{align}
% \bra\mathcal{O}\ket_0=\frac{\bra\mathcal{O}\ket}{Z(N,N_f)}
%\label{eq:normalVEV}
%\end{align}

As in the case of ABJM theory \cite{KMSS,HHMO},
we can study the Wilson loop VEV
of $N_f$ matrix model
from the Fermi gas approach.
As discussed in \cite{HHMO}, 
using the relation
\begin{align}
 \sum_{N=0}^\infty \ka^N\Biggl\bra\prod_{i=1}^N(1+\varepsilon e^{mx_i})\Biggr\ket
=\text{Det}\Bigl(1+\ka\rho(1+\varepsilon e^{mx})\Bigr)
\end{align}
and picking up the $\mathcal{O}(\varepsilon)$ term on both sides,
we find that the grand canonical VEV of winding Wilson loop is written as
\begin{align}
\begin{aligned}
 W_m(\mu,N_f)=\sum_{N=1}^\infty \ka^N
W_m(N,N_f)
=\mathrm{Det}(1+\ka\rho)\Tr\left(\frac{\ka\rho}{1+\ka\rho}e^{mx}\right). 
\end{aligned}
\end{align}
In the following, we will mainly consider the
grand canonical VEV of winding Wilson loops
normalized by
the grand partition function
\begin{align}
\frac{W_m(\mu,N_f)}{\Xi(\mu,N_f)}=\Tr\left(\frac{\ka\rho}{1+\ka\rho}e^{mx}\right).
\label{eq:grand-vev}
\end{align}

\subsection{Computation of trace}

To compute the grand canonical VEV of Wilson loop \eqref{eq:grand-vev},
we have to evaluate the trace $\Tr(\rho^\ell e^{mx})$
\begin{align}
 \frac{W_m(\mu,N_f)}{\Xi(\mu,N_f)}= \sum_{\ell=1}^\infty (-1)^{\ell-1}\ka^{\ell}\Tr(\rho^\ell e^{mx}).
\label{eq:ka-series}
\end{align}
This can be done systematically using the Tracy-Widom lemma \cite{TW}.
Noticing that $\rho$ is written as
\begin{align}
\begin{aligned}
 \rho(x,y)&=\frac{E(x)E(y)}{M(x)+M(y)},\\
M(x)=e^x &,\qquad
E(x)=\rt{\frac{1}{2\pi} \frac{e^x}{(2\cosh\frac{x}{2})^{N_f}}},
\end{aligned}
\label{eq:rho-tw}
\end{align}
one can show the $\ell^{\text{th}}$ power of $\rho$ is given by \cite{TW}
\begin{align}
 \rho^{\ell}(x,y)=\frac{E(x)E(y)}{M(x)+(-1)^{\ell-1}M(y)}\sum_{j=0}^{\ell-1}
(-1)^j\psi_j(x)\psi_{\ell-1-j}(y),
\end{align}
where the functions $\psi_j(x)~(j=0,1,\cdots)$ are determined
recursively starting from $\psi_0(x)=1$
\begin{align}
\begin{aligned}
\psi_j(x)&=\frac{1}{E(x)}\int dy\rho(x,y)E(y)\psi_{j-1}(y).
\end{aligned}
\end{align}
When $N_f$ is an integer, 
one can compute the exact values of VEV by closing the contour and 
picking up the residue of poles, as in the case of ABJM theory
\cite{Okuyama:2011su,Hatsuda:2012hm,Putrov:2012zi,Hatsuda:2012dt}.
In appendix \ref{app:exact},
we list some examples of the exact values of Wilson loop VEVs.

As explained in \cite{Okuyama:2016deu}, we can also compute
the sequence of functions
$\{\psi_j\}$ numerically 
with high precision by 
discrete Fourier transformations\footnote{We computed
the Wilson loop VEVs numerically using a \texttt{Mathematica}
program originally written by Yasuyuki Hatsuda.
We would like to thank him for sharing his program with us.}.
\begin{figure}[thb]
\centering\includegraphics[width=8cm]{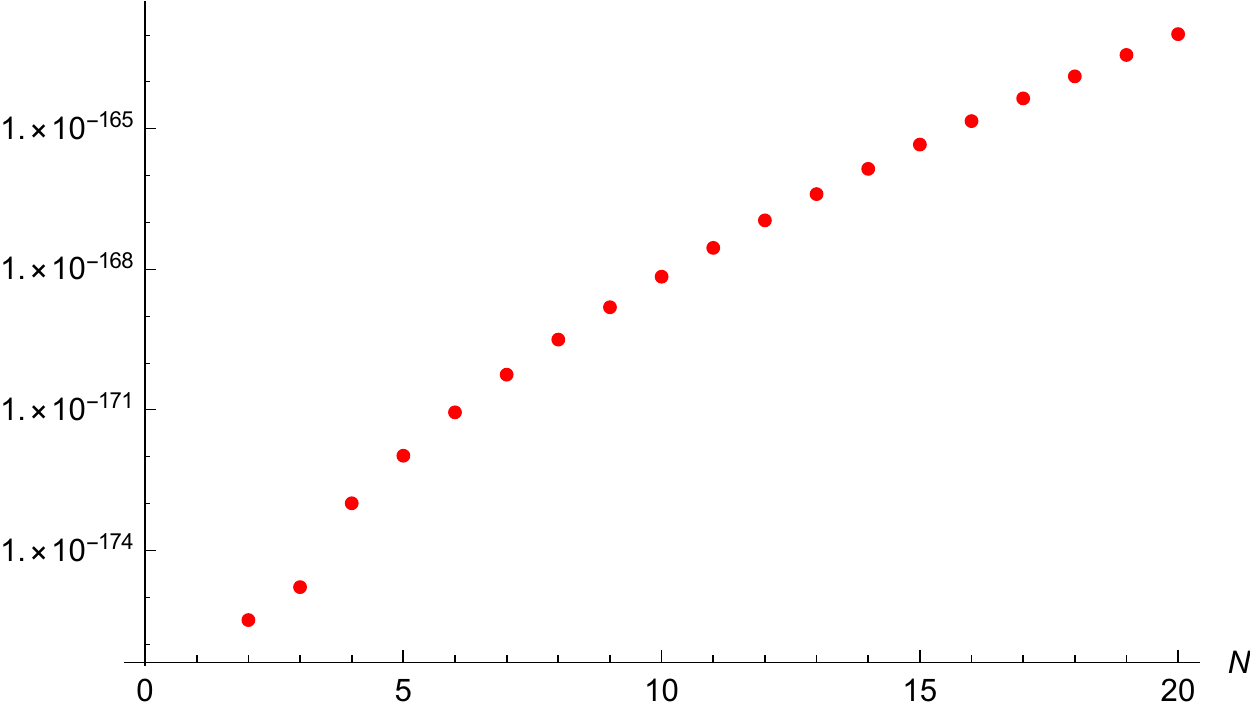}
  \caption{
Plot of the numerical error
$e_N$ \eqref{eq:def-eN} of the 
fundamental Wilson loop 
for $N_f=6$.
}
  \label{fig:error}
\end{figure}
We can estimate the
accuracy of our numerics by comparing
with the exact values in appendix \ref{app:exact}.
As an example, let us consider
the Wilson loop VEV in the fundamental representation
(or winding number $m=1$) for $N_f=6$. 
We have computed the exact values 
of $W_{\tableau{1}}(N,N_f=6)$\footnote{Since the
fundamental representation corresponds to the winding number $m=1$,
we will often use the notation $W_1=W_{\tableau{1}}$.
}
up to $N=20$, and in Fig.~\ref{fig:error}
we plotted the relative error 
 \begin{align}
  e_N=\left|\frac{W_{\tableau{1}}(N,6)_{\text{numerical}}}{W_{\tableau{1}}(N,6)_{\text{exact}}}-1\right|,
\label{eq:def-eN}
 \end{align}
between the exact and the numerical values.
As we can see from Fig.~\ref{fig:error}, 
our numerics has a high precision with an extremely small error
\begin{align}
 e_N < 1.1\times 10^{-163}\quad (\text{for}~~N\leq 20).
\label{eq:eN}
\end{align}
On the other hand, the instanton factors \eqref{eq:order-inst} for $N_f=6,N=20,$ are
given by 
\begin{align}
\begin{aligned}
 \text{worldsheet~1-instanton}&:~e^{-2\pi\rt{2N/N_f}}\approx 9.0\times 10^{-8},\\
\text{membrane~1-instanton}&:~e^{-2\pi\rt{N_fN/2}}\approx 7.3\times 10^{-22}.
\end{aligned}
\label{eq:ws-num}
\end{align}
From \eqref{eq:eN} and \eqref{eq:ws-num}, one can see that
our numerical computation has enough accuracy to study instanton corrections
to the Wilson loop VEVs.
Also, from \eqref{eq:ws-num}
we observe that the membrane instanton correction is highly suppressed compared to the
worldsheet instanton correction.
This is a generic phenomenon in the convergence region \eqref{eq:Nf-cond}.
Thus it is difficult to study membrane instanton corrections to the Wilson loops numerically;
in this paper we will mainly consider the worldsheet instanton corrections to the Wilson loop
VEVs.

\subsubsection{Odd winding number}
It turns out that the grand canonical VEV
in \eqref{eq:grand-vev}
for odd $m$ can be rewritten 
in a simpler form.
Let us first consider the fundamental representation.
We notice that Wilson loop factor $e^{mx}$ for $m=1$ reduces to $M(x)$
in \eqref{eq:rho-tw}.
Then the trace $\Tr (\rho^\ell M)$ can be simplified 
using the formal operator relation
\begin{align}
 M\rho+\rho M=|E\ket\bra E|.
\end{align}
Then we have
\begin{align}
 \begin{aligned}
  \Tr (\rho^\ell M)=\Tr(\rho^{\ell-1}\rho M)
=\Tr(\rho^{\ell-1}|E\ket\bra E|)-\Tr(\rho^{\ell-1}M\rho).
 \end{aligned}
\label{eq:m1-tr}
\end{align}
Using the cyclicity of trace, we find
\begin{align}
 \Tr (\rho^\ell M)=\hf \bra E|\rho^{\ell-1}|E\ket.
\end{align}
Finally, we find that the grand canonical VEV of fundamental Wilson loop is written as
\begin{align}
 \frac{W_{\tableau{1}}(\mu,N_f)}{\Xi(\mu,N_f)}
=\Tr\left(\frac{\ka\rho}{1+\ka\rho}M\right)
=\hf\bra E|\frac{\ka}{1+\ka\rho}|E\ket.
\end{align}
This is reminiscent of the grand canonical VEV of
1/2 BPS Wilson loops in ABJM theory \cite{HHMO}.
More generally, for odd $m$ one can show that
\begin{align}
\frac{W_m(\mu,N_f)}{\Xi(\mu,N_f)}
=\Tr\left(\frac{\ka\rho}{1+\ka\rho}M^m\right)
 =\hf\sum_{j=0}^{m-1}(-1)^j\bra E|M^j\frac{\ka}{1+\ka\rho}M^{m-1-j}|E\ket,
\end{align}
in a similar manner as \eqref{eq:m1-tr}.
On the other hand, we could not find a similar expression for
even winding numbers.

\section{Perturbative part}\label{sec:pert}
In this section, we consider the large $\mu$ expansion of the grand canonical VEV \eqref{eq:grand-vev}
of winding Wilson loops. 
It is useful to introduce the {\it modified version} 
of the grand canonical VEV $\h{W}_m(\mu,N_f)$  by
\begin{align}
W_m(\mu,N_f)= \sum_{n\in\mathbb{Z}}e^{J(\mu+2\pi\ri n,N_f)}\h{W}_m(\mu+2\pi \ri n,N_f).
\end{align}
As in the case of modified grand potential,
$\h{W}_m(\mu,N_f)$
is written as a sum of 
perturbative part and the exponentially suppressed
instanton corrections.

\begin{figure}[tb]
\centering
\includegraphics[width=10cm]{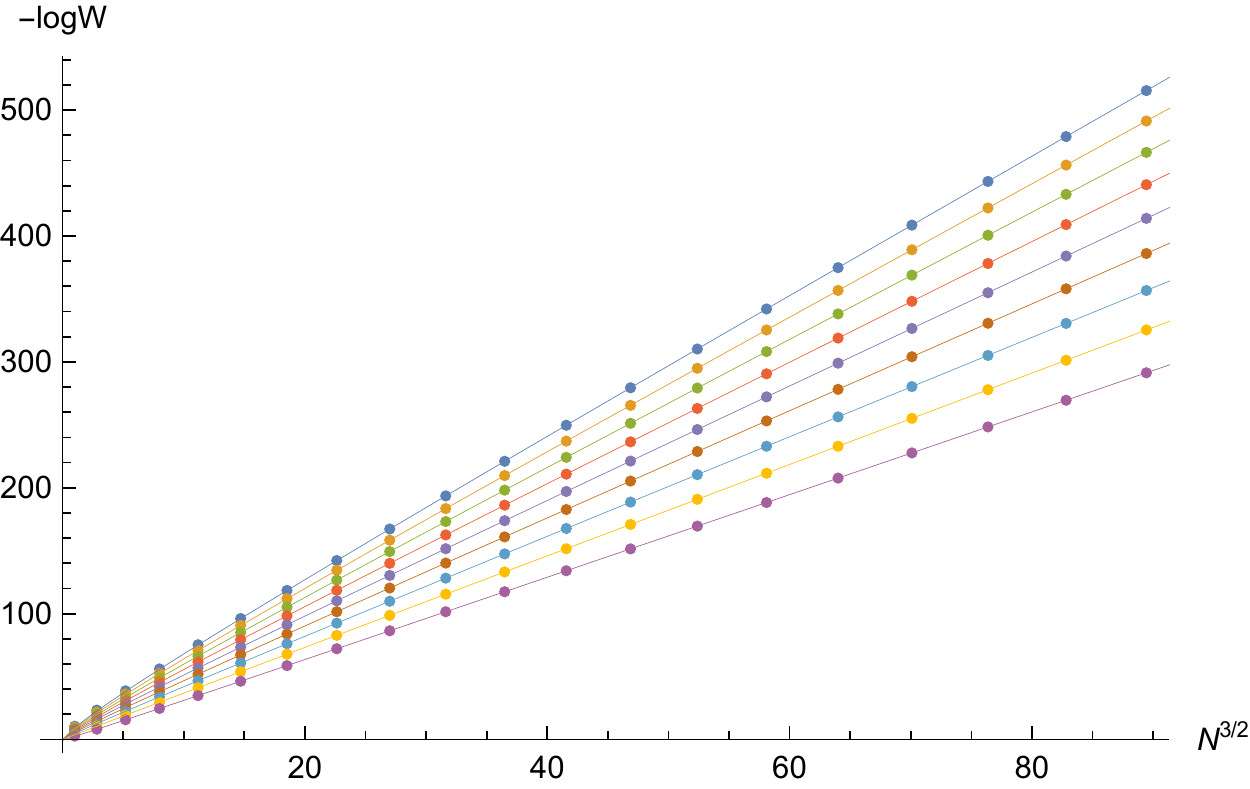}
  \caption{
We show the plot 
of $-\log W_2(N,N_f)$
for $N_f=5,6,\cdots,13$.
Note that the horizontal axis is $N^{3/2}$, and
$N_f$ increases from the bottom curve $(N_f=5)$
to the top curve $(N_f=13)$.
The dots are the numerical values of the Wilson loop VEV while
the solid curves represent the perturbative part given by
the Airy function \eqref{eq:Ai-pert}.
}
  \label{fig:Zpert}
\end{figure}

We find that the leading term (perturbative part)
of $\h{W}_m(\mu,N_f)$
in the large $\mu$ limit is given by
\begin{align}
 \h{W}_m^{\text{pert}}(\mu,N_f)= c_m(N_f)e^{\frac{2m\mu}{N_f}}.
\label{eq:Wm-scale}
\end{align}
The coefficient $c_m(N_f)$ in \eqref{eq:Wm-scale}
can be determined as follows.
As in the case of partition function \eqref{eq:mu-int},
the canonical picture and the grand canonical picture
of Wilson loop VEV
are related by the integral transformation
\begin{align}
 W_m(N,N_f)=\int_{\mathcal{C}}\frac{d\mu}{2\pi\ri}
e^{J(\mu,N_f)}\h{W}_m(\mu,N_f).
\label{eq:intC}
\end{align}
Then expanding the integrand of \eqref{eq:intC}
\begin{align}
 e^{J(\mu,N_f)}\h{W}_m(\mu,N_f)
=e^{J^{\text{pert}}(\mu,N_f)+\frac{2m\mu}{N_f}}\sum_{j,w}a_{j,w}\mu^je^{-w\mu},
\label{eq:Jexp-ajw}
\end{align}
the canonical VEV is written as
a sum of Airy function and its derivatives
\begin{align}
 W_m(N,N_f)=e^{A}C^{-\frac{1}{3}}\sum_{j,w}a_{j,w}
(-\del_N)^j\text{Ai}\Biggl[C^{-\frac{1}{3}}\Bigl(N-B-\frac{2m}{N_f}+w\Bigr)\Biggr].
\label{eq:Ai-exp}
\end{align}
In particular, the perturbative part of canonical VEV is given by
\begin{align}
 W_m^{\text{pert}}(N,N_f)=c_m(N_f)e^{A}C^{-\frac{1}{3}}\text{Ai}\Biggl[C^{-\frac{1}{3}}\Bigl(N-B-\frac{2m}{N_f}\Bigr)\Biggr].
\label{eq:Ai-pert}
\end{align}
By fitting the numerical value of $W_m(N,N_f)$ with the expression \eqref{eq:Ai-pert},
we can determine the coefficient $c_m(N_f)$.
In this way, we find
that the coefficient in the perturbative part 
of $\h{W}_m(\mu,N_f)$ in \eqref{eq:Wm-scale}
is given by
\begin{align}
\begin{aligned}
 c_1(N_f)&=\frac{1}{4\sin\frac{2\pi}{N_f}},\quad
&c_3(N_f)&=\frac{\sin^2\frac{\pi}{N_f}}{4\sin\frac{2\pi}{N_f}\sin\frac{4\pi}{N_f}\sin\frac{6\pi}{N_f}},\\
c_2(N_f) &=\frac{1}{2N_f\sin\frac{2\pi}{N_f}\sin\frac{4\pi}{N_f}},\quad
&c_4(N_f)&=\frac{\sin^2\frac{2\pi}{N_f}}{2N_f\sin\frac{2\pi}{N_f}\sin\frac{4\pi}{N_f}
\sin\frac{6\pi}{N_f}\sin\frac{8\pi}{N_f}},
\end{aligned}
\label{eq:c1-4}
\end{align}
for winding numbers $m=1,\cdots,4$.
For general winding number $m$, we conjecture
\begin{align}
c_m(N_f)=
\left\{
\begin{aligned}
&\frac{\prod_{j=1}^{\frac{m-1}{2}}\sin^2\frac{(2j-1)\pi}{N_f}}
{4\prod_{n=1}^m\sin\frac{2\pi n}{N_f}}&\quad&(\text{odd}~m),\\
&\frac{\prod_{j=1}^{\frac{m}{2}-1}\sin^2\frac{2j\pi}{N_f}}{2N_f\prod_{n=1}^m\sin\frac{2\pi n}{N_f}}
&\quad&(\text{even}~m).
\end{aligned}\right.
\label{eq:cm-coef}
\end{align}
We have checked this behavior numerically for  $m=1,\cdots,8$.
In Fig.~\ref{fig:Zpert}, we show the plot 
of Wilson loop VEV with  $m=2$  as an example.
As we can see from Fig.~\ref{fig:Zpert},
the Airy function in \eqref{eq:Ai-pert} exhibits a nice agreement with the numerical value
of $W_2(N,N_f)$,
if we use the correct coefficient $c_2(N_f)$ in \eqref{eq:c1-4}.
We have also confirmed a similar agreement 
between $W_m(N,N_f)$ and the Airy function \eqref{eq:Ai-pert}
for $m=1,\cdots,8$ with the coefficient
$c_m(N_f)$ in \eqref{eq:cm-coef}.

\section{Instanton corrections}\label{sec:inst}
We can continue the numerical fitting beyond the perturbative part
and fix the instanton coefficients in 
the expansion \eqref{eq:Jexp-ajw} and \eqref{eq:Ai-exp}.
As we will see below, we determine the first
few worldsheet instanton corrections to
$\h{W}_m(\mu,N_f)$ for $m=1,2,3$ in a closed form as a function of $N_f$. 
We conjecture that there is no ``pure'' membrane instanton corrections
to $\h{W}_m(\mu,N_f)$ except for the contributions of bound states.

\subsection{Fundamental representation}

Let us first consider the fundamental representation.
We find that the worldsheet instanton corrections
are given by
\begin{align}
 \begin{aligned}
\h{W}_{\tableau{1}}(\mu,N_f)&=
-\frac{N_f}{4}+e^{\frac{2\mu}{N_f}}\Biggl[
\frac{1}{4\sin\frac{2\pi}{N_f}}
+\frac{4\mu+N_f}{4\pi}e^{-\frac{4\mu}{N_f}}\\
&+\Biggl(\frac{\sin\frac{2\pi}{N_f}(4\mu+N_f)^2}{8\pi^2}
-\frac{3\cos\frac{2\pi}{N_f}(8\mu+N_f)}{8\pi}
+\frac{\cos^2\frac{2\pi}{N_f}}{\sin\frac{2\pi}{N_f}}\Biggr)e^{-\frac{8\mu}{N_f}}\\
&+\Biggl(\frac{(\sin^2\frac{2\pi}{N_f}+\sin^2\frac{4\pi}{N_f})(4\mu+N_f)^3}{24\pi^3}
-\frac{3\sin\frac{4\pi}{N_f}\sin\frac{6\pi}{N_f}}{\sin\frac{2\pi}{N_f}}\frac{(4\mu+N_f)(8\mu+N_f)}{16\pi^2}\\
&\qquad+\frac{5\cos\frac{4\pi}{N_f}\sin\frac{6\pi}{N_f}}{9\sin\frac{2\pi}{N_f}}\frac{12\mu+N_f}{\pi}
+(\cos^2\frac{2\pi}{N_f}+\cos^2\frac{4\pi}{N_f})\frac{4\mu+N_f}{\pi}\\
&\qquad-\frac{2(\sin\frac{6\pi}{N_f}+\cos\frac{6\pi}{N_f}\sin\frac{8\pi}{N_f})}{\sin\frac{2\pi}{N_f}\sin\frac{4\pi}{N_f}}\Biggr)
e^{-\frac{12\mu}{N_f}}+\mathcal{O}\Bigl(e^{-\frac{16\mu}{N_f}}\Bigr)\Biggr].
 \end{aligned}
\label{eq:fund-inst-fermi}
\end{align}

\begin{figure}[tb]
\centering
\subcaptionbox{$N_f=7$\label{fig:dNf7}}{\includegraphics[width=4.5cm]{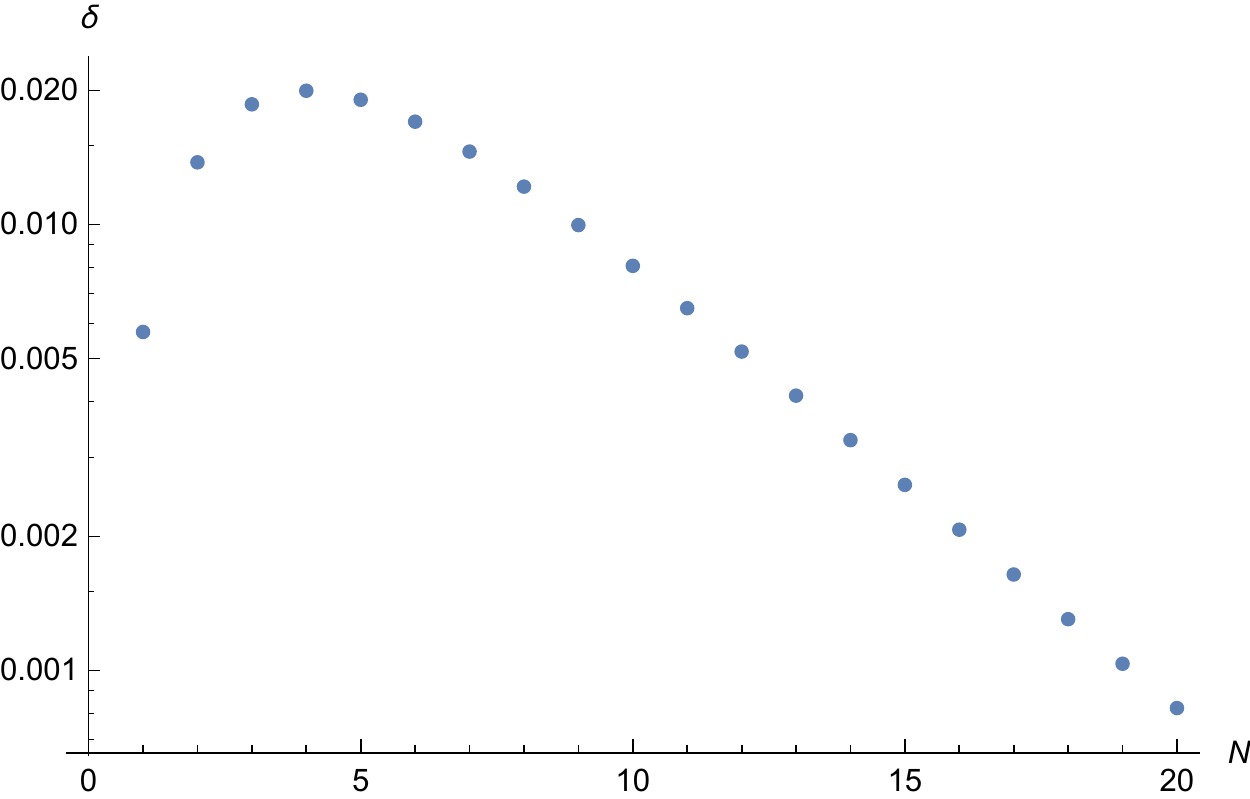}}
\hskip2mm
\subcaptionbox{$N_f=8$\label{fig:dNf8}}{\includegraphics[width=4.5cm]{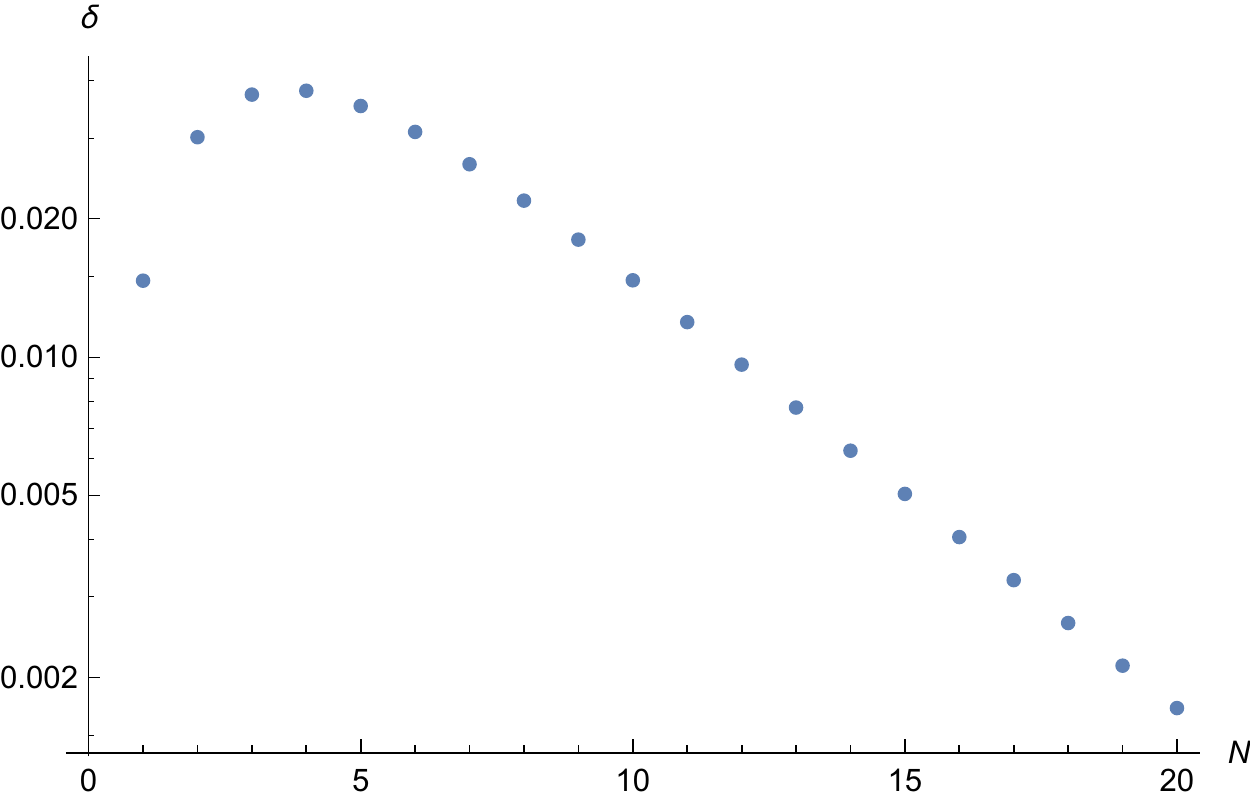}}
\hskip2mm
\subcaptionbox{$N_f=11$\label{fig:dNf11}}{\includegraphics[width=4.5cm]{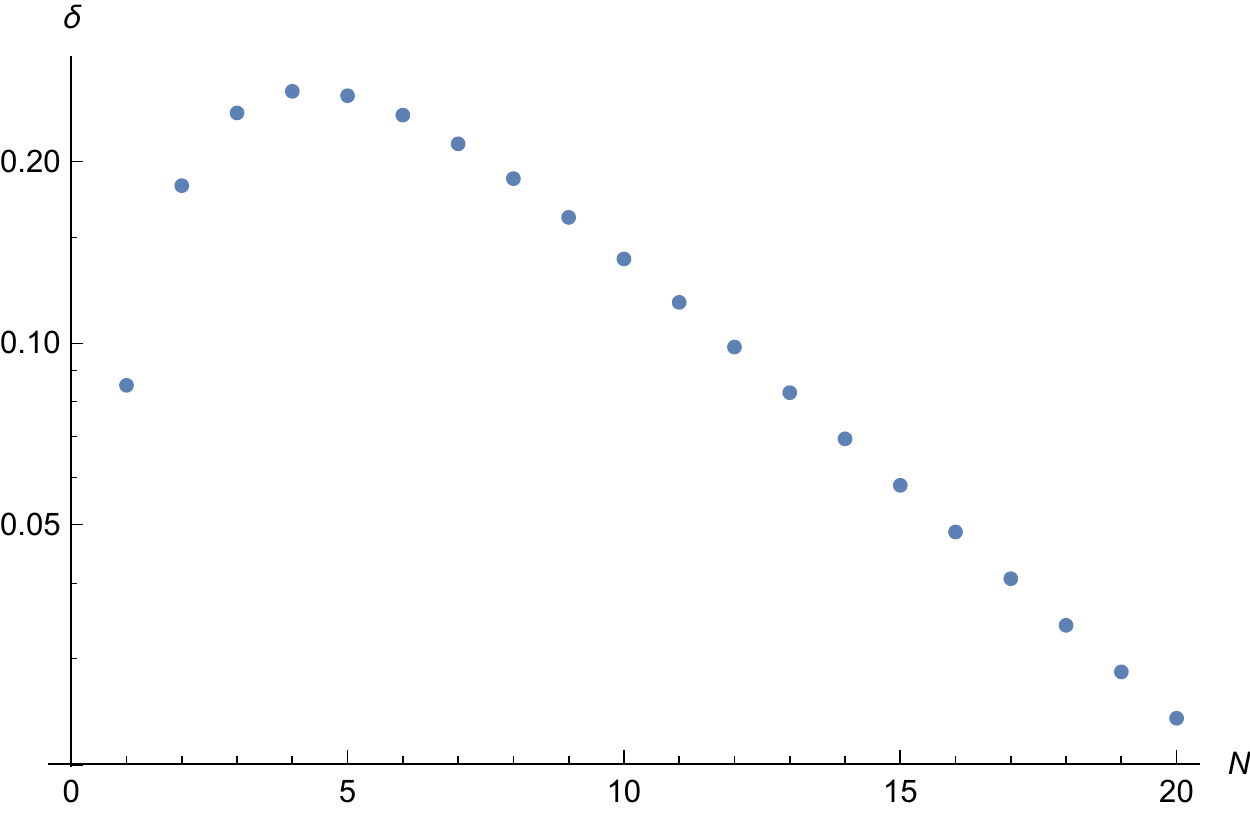}}
  \caption{
We show the plot 
of $\cob$ for $N_f=7,8,11$.
Note that the vertical axis is log scale.
}
  \label{fig:d-fund}
\end{figure}

In Fig.~\ref{fig:d-fund} we plot the quantity
\begin{align}
 \cob=\frac{W_{\tableau{1}}(N,N_f)-W_{\tableau{1}}^{\text{pert}}(N,N_f)-W_{\tableau{1}}^{\text{inst}}(N,N_f)}{W_{\tableau{1}}^{\text{pert}}(N,N_f)}e^{\frac{12\mu_*}{N_f}}
\label{eq:def-cob}
\end{align}
where $W_{\tableau{1}}^{\text{pert}}(N,N_f)$ in \eqref{eq:def-cob}
is the perturbative part given by the Airy function \eqref{eq:Ai-pert} 
and $\mu_*$
is the saddle point value of the chemical potential 
in \eqref{eq:mu-saddle}, and
$W_{\tableau{1}}^{\text{inst}}(N,N_f)$ is the instanton correction
in the canonical picture up to worldsheet 3-instantons, 
obtained from the grand canonical picture 
\eqref{eq:fund-inst-fermi} using 
\eqref{eq:Jexp-ajw} and \eqref{eq:Ai-exp}.
If we have subtracted worldsheet instantons correctly
in \eqref{eq:def-cob}, $\cob$ should
decay exponentially as $N$ becomes large.
Indeed, in Fig. \ref{fig:d-fund}
we find that $\cob$
decays exponentially for $N_f=7,8,11$, as expected.
We have also checked a similar behavior of $\cob$ for other values of $N_f$. 
This confirms the correctness of the instanton corrections in \eqref{eq:fund-inst-fermi}.

In 
\eqref{eq:fund-inst-fermi},
we observe that the worldsheet 1-instanton and 2-instanton 
have no poles in the convergence region $N_f>2$ \eqref{eq:Nf-cond}.
This suggests that there is no ``pure'' membrane instanton corrections
as in the case of 1/2 BPS Wilson loops in ABJM theory \cite{HHMO},
since there is no need for the membrane instantons to appear to cancel
the poles.
On the other hand,
the worldsheet 3-instanton has a pole at $N_f=4$.
We conjecture that this pole is canceled by the bound state
of order $e^{-2\mu-4\mu/N_f}$.
It would be interesting to determine the coefficient of this bound state
contribution as a function of $N_f$.

Also, we observe that 
the grand canonical VEV \eqref{eq:fund-inst-fermi}
has two pieces which scale differently in the large $\mu$ limit:
the constant term $-N_f/4$ and the remaining part
whose leading term gives rise to the perturbative part \eqref{eq:Wm-scale}.
One can translate this decomposition into the canonical picture
\begin{align}
 W_{\tableau{1}}(N,N_f)=-\frac{N_f}{4}Z(N,N_f)+\til{W}_{\tableau{1}}(N,N_f),
\label{eq:fund-decomp}
\end{align}
where the second term behaves in the large $N$ limit
as
\begin{align}
 \frac{\til{W}_{\tableau{1}}(N,N_f)}{Z(N,N_f)}\approx c_1(N_f)e^{\frac{2\mu_*}{N_f}}.
\end{align}
In section~\ref{sec:tHooft}, we will show that this decomposition \eqref{eq:fund-decomp} is consistent with
the genus-zero result in \cite{Grassi:2014vwa}.

\subsection{Winding number $m=2,3$}
Next we consider the worldsheet instanton 
corrections
to the winding Wilson loops $\h{W}_m(\mu,N_f)$
for winding numbers $m=2,3$.

\paragraph{Winding number $\boldsymbol{m=2}$}
For  $m=2$ we find
\begin{align}
\begin{aligned}
\h{W}_2(\mu,N_f)=&\frac{2e^{\frac{4\mu}{N_f}}}{N_f\sin\frac{2\pi}{N_f}}\Biggl[\frac{1}{4\sin\frac{4\pi}{N_f}}
+\frac{\cos\frac{2\pi}{N_f}(4\mu+N_f)}{4\pi}e^{-\frac{4\mu}{N_f}} \\
&+\Biggl(\frac{\sin\frac{4\pi}{N_f}(4\mu+N_f)^2}{8\pi^2}
-\frac{3\cos\frac{4\pi}{N_f}(8\mu+N_f)}{8\pi}+\frac{\cos\frac{2\pi}{N_f}\cos\frac{4\pi}{N_f}}
{\sin\frac{2\pi}{N_f}}\Biggr)e^{-\frac{8\mu}{N_f}}
+\mathcal{O}(e^{-\frac{12\mu}{N_f}})\Biggl]\\
&-C\mu^2+B-\frac{8\mu}{\pi N_f\sin\frac{2\pi}{N_f}}e^{-\frac{4\mu}{N_f}}
+\mathcal{O}(e^{-\frac{8\mu}{N_f}}).
\end{aligned}
\label{eq:w2-inst}
\end{align}
We observe that the last line of \eqref{eq:w2-inst} is related to the derivative
of the modified grand potential
\begin{align}
 -\del_\mu J(\mu,N_f)+2B,
\end{align}
where $B$ is the coefficient in the perturbative part \eqref{eq:abc}.
As in the case of fundamental representation in the previous subsection,
$\h{W}_2(\mu,N_f)$ consists of two parts with different scaling behavior in the large
$\mu$ limit,
which  implies that the Wilson loop VEV $W_2(N,N_f)$ in the canonical picture
can be decomposed as
\begin{align}
 W_2(N,N_f)=(-N+2B)Z(N,N_f)+\til{W}_2(N,N_f),
\label{eq:decom-w2}
\end{align}
where the second term in \eqref{eq:decom-w2}
behaves in the large $N$ limit as
\begin{align}
 \frac{\til{W}_2(N,N_f)}{Z(N,N_f)}\approx c_2(N_f)e^{\frac{4\mu_*}{N_f}}.
\end{align}

\paragraph{Winding number $\boldsymbol{m=3}$}
For  $m=3$ we find
\begin{align}
\begin{aligned}
\h{W}_3(\mu,N_f)
 =&\frac{\sin^2\frac{\pi}{N_f}}{\sin\frac{2\pi}{N_f}\sin\frac{4\pi}{N_f}}
e^{\frac{6\mu}{N_f}}\Biggl[\frac{1}{4\sin\frac{6\pi}{N_f}}-\frac{\sin\frac{3\pi}{N_f}}{\sin\frac{\pi}{N_f}}\frac{(4\mu+N_f)}{4\pi}e^{-\frac{4\mu}{N_f}}\\
&+\Biggl(\frac{\sin\frac{6\pi}{N_f}(4\mu+N_f)^2}{8\pi^2}
-\frac{3\cos\frac{2\pi}{N_f}\sin^2\frac{3\pi}{N_f}}{\sin^2\frac{\pi}{N_f}}\frac{(8\mu+N_f)}{8\pi}\\
&+\frac{\sin\frac{3\pi}{N_f}\cos\frac{2\pi}{N_f}}{\sin\frac{\pi}{N_f}\sin\frac{2\pi}{N_f}}\Bigl(\cos\frac{4\pi}{N_f}-\cos\frac{2\pi}{N_f}-1\Bigr)\Biggr)e^{-\frac{8\mu}{N_f}}
+\mathcal{O}(e^{-\frac{12\mu}{N_f}})\Biggr]\\
&-\frac{N_f}{4}+ 3e^{\frac{2\mu}{N_f}}\Biggl[\frac{1}{4\sin\frac{2\pi}{N_f}}
+\frac{4\mu+N_f}{4\pi}e^{-\frac{4\mu}{N_f}}+\mathcal{O}(e^{-\frac{8\mu}{N_f}})\Biggr].
\end{aligned}
\label{eq:w3-inst}
\end{align}
In this case, we observe that $\h{W}_3(\mu,N_f)$
consists of three parts with different scalings in the large $\mu$ limit.
Noticing that the last term of \eqref{eq:w3-inst}
is proportional to $\til{W}_{\tableau{1}}$
in \eqref{eq:fund-inst-fermi},
we conjecture that the canonical VEV $W_3(N,N_f)$ 
for winding number $m=3$ is decomposed as
\begin{align}
\begin{aligned}
 W_3(N,N_f)=-\frac{N_f}{4}Z(N,N_f)+3\til{W}_{\tableau{1}}(N,N_f)+\til{W}_3(N,N_f),
\end{aligned}
\label{eq:decom-w3}
\end{align}
where the last term scales as
\begin{align}
 \frac{\til{W}_3(N,N_f)}{Z(N,N_f)}\approx c_3(N_f)e^{\frac{6\mu_*}{N_f}}.
\end{align}

We have confirmed our result of instanton corrections \eqref{eq:w2-inst} and \eqref{eq:w3-inst}
for $m=2,3$
by performing a similar test as in Fig. \ref{fig:d-fund}.
We have checked a correct exponential decay of the quantity $\delta$ for various $N_f$'s
in the convergence region \eqref{eq:Nf-cond}.

\section{WKB expansion}\label{sec:wkb}
In this section, we will
consider the WKB expansion (small $\hbar$ expansion)
of spectral trace $\Tr(\rho^s e^x)$ and
try to reproduce the perturbative part of fundamental representation.
Our starting point is the Mellin-Barnes representation of the
grand canonical VEV \cite{Hatsuda:2015oaa,Marino:2016new}
\begin{align}
\h{W}_{\tableau{1}}(\mu,N_f)=\int_{c-\ri\infty}^{c+\ri\infty}\frac{ds}{2\pi\ri}\frac{\pi}{\sin\pi s}\Tr(\rho^se^x)e^{s\mu},
\label{eq:MB-int}
\end{align}
where $c$ is a positive constant in the region $2/N_f<c<1$.
By picking up the poles at $s=\ell\in\mathbb{N}$
we recover the small $\ka$ expansion \eqref{eq:ka-series}.
On the other hand, closing the contour
in the direction $\text{Re}(s)<c$ we can find the 
large $\mu$ expansion of $\h{W}_{\tableau{1}}(\mu,N_f)$.

In the quantum mechanical description of density matrix \eqref{eq:rho-QM}, 
the Planck constant is fixed to $\hbar=2\pi$ \eqref{eq:hbar-2pi}.
However, one can formally introduce
the parameter $\hbar$ in the canonical commutation relation 
$[\h{x},\h{p}]=\ri\hbar$
and perform the WKB expansion of the spectral trace $\Tr(\rho^se^x)$.
Finally we set $\hbar=2\pi$ at the end of computation.
This procedure was successfully applied to several
examples \cite{Assel:2015hsa,Okuyama:2015auc}.

At the zero-th order of WKB expansion,
the spectral trace is given by the classical phase space integral
\begin{align}
Z_0(s)= \int\frac{dxdp}{2\pi\hbar}
\left[\frac{1}{2\cosh\frac{p}{2}}\frac{1}{(2\cosh\frac{x}{2})^{N_f}}\right]^se^{x}=
\frac{\Ga\big(\frac{s}{2}\big)^2\Ga\Big(\frac{N_fs}{2}+1\Big)\Ga\Big(\frac{N_fs}{2}-1\Big)}{2\pi\hbar\Ga(s)\Ga(N_fs)},
\label{eq:Z0}
\end{align}
and the higher order corrections can be systematically computed by using 
the Wigner transformation of operator $\hat{\rho}^se^{\h{x}}$
\cite{Hatsuda:2015lpa,Okuyama:2015auc,Okuyama:2016xke,Okuyama:2016deu}. In this way, we find the WKB expansion of spectral trace as 
\begin{align}
\begin{aligned}
 \Tr(\rho^sx^x)=Z_0(s)D(s),
\end{aligned}
\end{align}
where $D(s)$ is a formal power series in $\hbar$
\begin{align}
 D(s)=1+\sum_{n=1}^\infty \hbar^{2n}D_n(s).
\end{align}
We find that the $n^{\text{th}}$ order term $D_n(s)$
has the following structure:
\begin{align}
 D_n(s)=\frac{(s-1)p_n(s)}{\prod_{j=1}^n(N_fs+2j-1)},
\end{align}
where $p_n(s)$ is a $(3n-1)^{\text{th}}$ order polynomial of $s$.
The first three terms are given by
\begin{align}
\begin{aligned}
 p_1(s)&=c_1(N_f^2 s^2-4 N_f s-8),\\
p_2(s)&=c_2\Biggl[N_f^4 s^5+\frac{1}{7} N_f^3 (3 N_f-32) s^4+\frac{8}{7} N_f^2
   \left(N_f^2-2 N_f-18\right) s^3\\
&\quad-\frac{16}{7} N_f \left(4 N_f^2+5
   N_f-24\right) s^2-\frac{128}{7} \left(3 N_f^2-N_f-9\right)
   s+\frac{384}{7}\Biggr],\\
p_3(s)&=c_3\Biggl[N_f^6 s^8+\frac{4}{31} N_f^5 (16 N_f-15) s^7+\frac{1}{31} N_f^4
   \left(129 N_f^2-176 N_f-1544\right) s^6\\
&\qquad+\frac{4}{31} N_f^3
   \left(18 N_f^3-129 N_f^2-944 N_f+288\right) s^5\\
&\qquad+\frac{8}{31}
   N_f^2 \left(16 N_f^4-44 N_f^3-1275 N_f^2-80 N_f+2640\right)
   s^4\\
&\qquad+\frac{64}{31} N_f \left(2 N_f^5-24 N_f^4-103 N_f^3-53 N_f^2+540
   N_f-72\right) s^3\\
&\qquad-\frac{128}{31} \left(12 N_f^5+120 N_f^4-32
   N_f^3-955 N_f^2+48 N_f+1080\right) s^2\\
&\qquad-\frac{1024}{31}
   \left(15 N_f^4-120 N_f^2+N_f+180\right) s-\frac{30720}{31}\Biggr].
\end{aligned}
\label{eq:pn(s)}
\end{align}
The coefficient $c_n$ 
of the highest order term $N_f^{2n}s^{3n-1}$
in $p_n(s)$ \eqref{eq:pn(s)}
is found to be
\begin{align}
 c_n=\frac{B_{2n}(1/2)}{2^{4n}(2n)!},
\end{align}
where $B_{2n}(1/2)$ denotes the  Bernoulli polynomial
$B_{2n}(z)$ evaluated at $z=1/2$. 
We have computed $D_n(s)$ up to $n_{\text{max}}=10$.

\begin{figure}[thb]
\centering
\subcaptionbox{plot of $D(2/N_f)$\label{fig:wkb-a}}{\includegraphics[width=7cm]{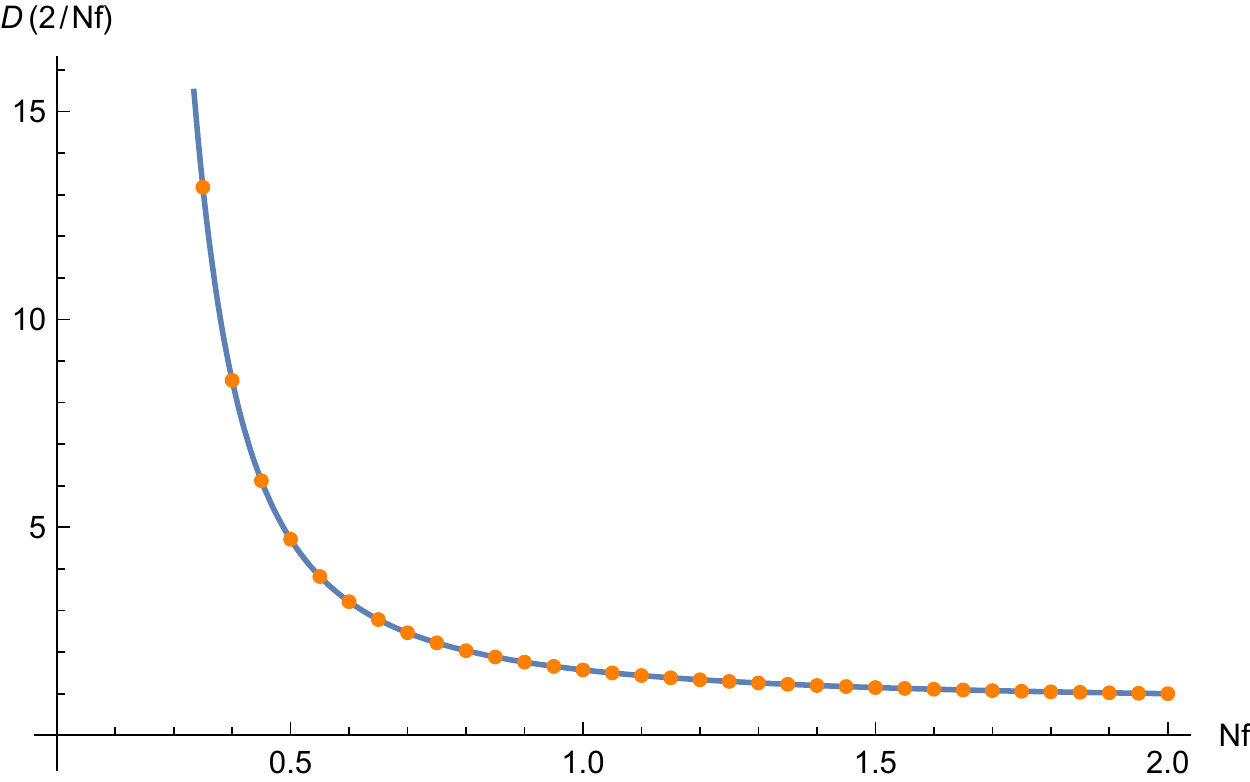}}
\hspace{5mm}
\subcaptionbox{plot of $D'(0)$\label{fig:wkb-b}}{\includegraphics[width=7cm]{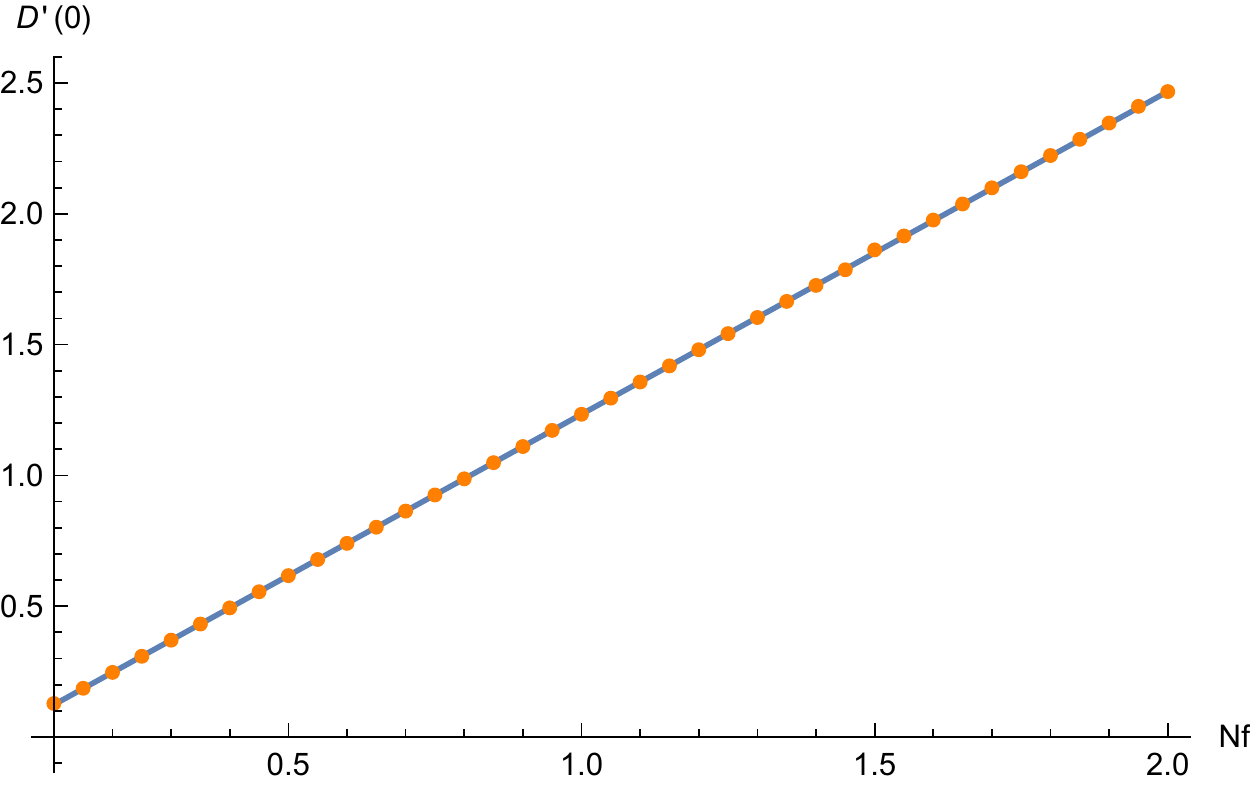}}
  \caption{
We show the plot 
of \subref{fig:wkb-a} $D(2/N_f)$ and \subref{fig:wkb-b} $D'(0)$,
against $N_f$.
The dots are the 
numerical values obtained from the Pad\'e approximation,
while
the solid curves represent 
the exact functions in \eqref{eq:Dpert-exact}
and \eqref{eq:Dconst-exact} for \subref{fig:wkb-a} and \subref{fig:wkb-b}, respectively.
}
  \label{fig:wkb}
\end{figure}

As mentioned above, 
the large $\mu$ expansion of grand canonical VEV
can be found by closing the contour of \eqref{eq:MB-int} in the direction $\text{Re(s)}<c$.
The perturbative part comes from the pole at $s=2/N_f$
of $Z_0(s)$ in \eqref{eq:Z0}
\begin{align}
 \text{Res}_{s=\frac{2}{N_f}}\left[\frac{\pi}{\sin\pi s}Z_0(s)D(s)e^{s\mu}\right]
=\frac{1}{N_f\hbar\sin\frac{2\pi}{N_f}}\frac{\Ga\bigl(\frac{1}{N_f}\bigr)^2}{\Ga\bigl(\frac{2}{N_f}\bigr)}D\Bigl(\frac{2}{N_f}\Bigr)e^{\frac{2\mu}{N_f}}.
\end{align}
In order to reproduce the coefficient $c_1(N_f)$ 
of the 
perturbative part of fundamental representation in \eqref{eq:c1-4},
we need to show that
\begin{align}
 D\Bigl(\frac{2}{N_f}\Bigr)=\frac{\pi N_f\Ga\bigl(\frac{2}{N_f}\bigr)}{2\Ga\bigl(\frac{1}{N_f}\bigr)^2}.
\label{eq:Dpert-exact}
\end{align}
Although we do not have an analytic proof of this relation,
we can check this numerically by
using the Pad\'{e} approximation
\begin{align}
 D(s)\approx 1+\sum_{n=1}^{n_{\text{max}}}\hbar^{2n}D_n(s)
\approx \frac{1+a_1\hbar^2+\cdots +a_{n_{\text{max}/2}}\hbar^{n_{\text{max}}}}
{1+b_1\hbar^2+\cdots +b_{n_{\text{max}/2}}\hbar^{n_{\text{max}}}},
\end{align}
 and set $\hbar=2\pi$ at the end.
As we can see from Fig.~\ref{fig:wkb-a},
the Pad\'{e} approximation
exhibits a nice agreement with the 
expected behavior in \eqref{eq:Dpert-exact}.

One can also repeat the same analysis for the constant part $-N_f/4$
in \eqref{eq:fund-inst-fermi},
which comes from the pole at $s=0$ in \eqref{eq:MB-int}
\begin{align}
 \text{Res}_{s=0}\left[\frac{\pi}{\sin\pi s}Z_0(s)D(s)e^{s\mu}\right]
=-\frac{2D(0)}{\pi\hbar}(2\mu+N_f)-\frac{4D'(0)}{\pi\hbar}.
\end{align}
From the first few terms of the expansion of $D(0)$,
one can easily guess the closed form of $D(0)$
\begin{align}
 D(0)=1-\frac{\hbar^2}{48}-\frac{\hbar^4}{11520}-\frac{\hbar^6}{1935360}+\cdots=
\frac{\hbar}{4}\cot\frac{\hbar}{4}.
\end{align}
After setting $\hbar=2\pi$, we find $D(0)=0$ as expected.
In order to reproduce the constant $-N_f/4$, we need
\begin{align}
 D'(0)=\frac{\pi^2N_f}{8}.
\label{eq:Dconst-exact}
\end{align}
Again, we can check this relation numerically using the 
Pad\'{e} approximation for $D'(0)$. 
We find a nice agreement
with the right hand side of \eqref{eq:Dconst-exact} (see Fig.~\ref{fig:wkb-b}).
 
We expect that the worldsheet $\ell$-instanton comes from
the pole at $s=(2-4\ell)/N_f$.
In principle, we can find the instanton coefficients
from the WKB analysis. 
However, it is difficult in practice
since both the classical part $Z_0(s)$
and the quantum corrections $D(s)$ should have poles at $s=(2-4\ell)/N_f$
in order to reproduce the Fermi gas result \eqref{eq:fund-inst-fermi}\footnote{A similar
problem occurs for the WKB analysis of the worldsheet instantons
in ABJM theory \cite{Hatsuda:2015oaa}.}.
This is different from the situation for the perturbative part and the constant term 
considered above, where the poles only come  from the classical part.
By the same reason, it is difficult to study the winding Wilson loop
$\h{W}_m(\mu,N_f)$ with 
$m\geq2$ from the WKB analysis.

\section{Planar solution of $N_f$ matrix model}\label{sec:planar}
In this section, we will summarize the planar solution of the $N_f$
matrix model.
This section is mostly a review of the result in
\cite{Grassi:2014vwa}, but we find that the 
the planar resolvent in \cite{Grassi:2014vwa}
can be vastly simplified.
Using this simplified expression of resolvent,
we will directly show  that 
the resolvent satisfies the planar loop equation
 without referring to the relation to the
$O(n)$ matrix model \cite{EK1,EK2,Suyama:2012uu}.

Let us consider the planar resolvent
of $N_f$ matrix model in the 't Hooft limit \eqref{eq:thooft-limit}
\begin{align}
 \om(z)=\frac{1}{Z(N,N_f)}
\Biggl\bra \frac{1}{N_f}\sum_{i=1}^N\frac{z+e^{x_i}}{z-e^{x_i}}\Biggr\ket,
\label{eq:om-def}
\end{align}
where we have normalized the VEV \eqref{eq:matrix-vev}
by the partition function.\footnote{We have change the notation of
resolvent from $v(z)$ in \cite{Grassi:2014vwa} to $\om(z)$,
in order to save the letter $v$ for the coordinate
on the torus $v\in \mathbb{C}/(\mathbb{Z}+\tau\mathbb{Z})$.}
In the 't Hooft limit \eqref{eq:thooft-limit},
the eigenvalue distribution becomes continuous. 
Since the integrand of the $N_f$ matrix model \eqref{eq:Zint} is an even function of
$x_i$, the eigenvalues are distributed symmetrically around the origin
$x_i\in[-T,T]$.
Noticing that the variable $z$ in \eqref{eq:om-def}
is related to the eigenvalue $x$ by $z=e^x$,
$\om(z)$ is expected to have a cut along $z\in[a,b]$ with
$a=e^{-T}, b=e^T$.

The resolvent $\om(z)$ should satisfy several
conditions.
First of all, $\om(z)$ should satisfy the loop equation which comes from the 
saddle point equation
of matrix integral \eqref{eq:Zint}
\begin{align}
 \om(z+\ri 0)+\om(z-\ri0)-2\om(-z)=V'(z)
\end{align}
along the cut $z\in[a,b]$. Here $V'(z)$ is given by
\begin{align}
 V'(z)=\frac{z-1}{z+1}.
\end{align}
Also, from the definition \eqref{eq:om-def} and the symmetry
of the eigenvalue distribution we find
\begin{align}
 \om(0)&=-\la,
\label{eq:om-la} \\
 \om(z^{-1})&=-\om(z).
\label{eq:om-inv}
\end{align}

\paragraph{Mapping to the torus $\mathbb{C}/(\mathbb{Z}+\tau\mathbb{Z})$}
To find the resolvent,
it is convenient to map $z$ to the variable $u$ by
a Jacobi elliptic function
\begin{align}
 z=a\,\text{sn}(u,k),
\label{eq:z-snu}
\end{align}
where the elliptic modulus $k$ is given by
\begin{align}
 k=\frac{a}{b}=a^2.
\end{align}
In what follows we will suppress the dependence of $k$.
We also need the derivative of $z$
\begin{align}
 \del_u z=a\,\text{cn}(u)\text{dn}(u)=a\rt{(1-a^2z^2)(1-a^{-2}z^2)}.
\label{eq:del-z}
\end{align}

Furthermore, the variable $u$ 
is related to the flat coordinate $v$
on the torus $\mathbb{C}/(\mathbb{Z}+\tau\mathbb{Z})$ by
\begin{align}
 v=\frac{u}{2K},
\end{align}
where $K=K(k^2)$ is the elliptic integral of the first kind
and the complex structure of the torus is given by
\begin{align}
\tau=\frac{\ri K'}{K}.
\end{align}
Then the coordinate $z$
and the end-point of the cut $a$
is written in terms of the Jacobi theta functions
\begin{align}
\begin{aligned}
 z&=\frac{\vartheta_1(v,\tau)}{\vartheta_4(v,\tau)},\\ 
a&=\frac{\vartheta_2(0,\tau)}{\vartheta_3(0,\tau)}.
\end{aligned}
\end{align}
We will use the variables $z,u$ and $v$ interchangeably.
One can show that $z(v)$ satisfies
\begin{align}
\begin{aligned}
&z(-v)=-z(v),\\
&z(v+1)=-z(v),\\
 &z(v+\tau)=z(v),\\
&z\Big(v+\frac{\tau}{2}\Big)=\frac{1}{z(v)},\\
&z\Big(v+\hf\Big)=a\rt{\frac{1-a^{-2}z(v)^2}{1-a^{2}z(v)^2}}
=\frac{\vartheta_2(v,\tau)}{\vartheta_3(v,\tau)},
\end{aligned}
\end{align}
and the various points 
on the $z$-plane are mapped to the points on the $v$-plane as
\begin{center}
\begin{tabular}{c|cccccc}
 $z$ & 0& $\infty$ & $a$ & $a^{-1}$ & $\pm1$ & $\pm\ri$ \\
\hline
$v$ & 0 & $\frac{\tau}{2}$ & $\hf$ & $\hf+\frac{\tau}{2}$ & $\pm\hf+\frac{\tau}{4}$
& $ \pm\frac{\tau}{4}$
\end{tabular} 
\end{center}

\paragraph{Conditions obeyed by the resolvent $\boldsymbol{\om(v)}$}
Here we will write down the conditions for the 
resolvent $\om(v)$ in terms of the variable $v$.
The loop equation becomes
\begin{align}
 \om\Bigl(\hf+v\Bigr)+\om\Bigl(\hf-v\Bigr)-2\om\Bigl(-\hf-v\Bigr)= V'\Bigl(\hf+v\Bigr),\quad
\text{for}~v\in[0,\frac{\tau}{2}] .
\end{align}
We also require that there are no cuts along $z\in[-b,-a]$ and
$z\in[-\infty,-b]\cup[b,\infty]$.
In terms of the variable $v$ these conditions become
\begin{align}
 \begin{aligned}
\om\Bigl(-\hf+v\Bigr)&=\om\Bigl(-\hf-v\Bigr),\quad
\text{for}~v\in[0,\frac{\tau}{2}] ,\\
\om\Bigl(\frac{\tau}{2}+v\Bigr)&=\om\Bigl(-\frac{\tau}{2}+v\Bigr)
,\quad
\text{for}~v\in[-\hf,\hf].
 \end{aligned}
\end{align}
Using the above conditions, 
$\om(v)$ can be extended to 
a function on the torus
obeying the following functional relations
\begin{align}
 &\om(v+1)+\om(v-1)-2\om(v)=V'(v+1),
\label{eq:om-loop}\\
&\om\Bigl(-\hf+v\Bigr)=\om\Bigl(-\hf-v\Bigr),
\label{eq:om-c1}\\
&\om(v+\tau)=\om(v).
\label{eq:om-c2}
\end{align}

\subsection{Planar resolvent}
The resolvent $\om(v)$
was obtained in \cite{Grassi:2014vwa}
by taking a limit of the solution of $O(n)$ matrix model
\cite{EK1,EK2,Suyama:2012uu}.
After massaging the expression in \cite{Grassi:2014vwa},
we find a very simple formula for the resolvent
\begin{align}
 \om(z)=\frac{\til{A}(z)^2}{2\pi^2}\frac{z^2+1}{z^2-1}
-\qu\frac{z+1}{z-1}+\h{\la}\,\frac{z^2-1}{z^2+1}.
\label{eq:om-z}
\end{align}
The function $\til{A}(z)$ in \eqref{eq:om-z}
is given by
\begin{align}
 \til{A}(z)=\frac{\pi}{2}+A(u)-K'\frac{\del_u z}{z(1+z^2)},
\label{eq:tAvsA}
\end{align}
where  $K'=K(1-k^2)$ and
$A(u)$ denotes the function introduced in \cite{Grassi:2014vwa}
\begin{align}
 A(u)=\frac{\pi u}{2K}+K'Z(u)+K'\del_u\log z,
\end{align}
with $Z(u)$ being the Jacobi zeta function
\begin{align}
 Z(u)=\del_u\log\vartheta_4\Bigl(\frac{u}{2K},\tau\Bigr).
\end{align}
In terms of the variable $v$,
we find that $\til{A}(v)$ in \eqref{eq:om-z}
is written as
\begin{align}
\begin{aligned}
 \til{A}(v)&=\frac{\pi}{2}+\pi v-\frac{\ri\tau}{2}
\del_v\log\vartheta_4(v,\tau)
-\frac{\ri\tau}{4}\del_v\log(1+z^2) \\
&=\frac{\pi}{2}+\pi v
-\frac{\ri\tau}{4}\del_v\log\Bigl[\vartheta_1(v,\tau)^2+\vartheta_4(v,\tau)^2\Bigr].
\end{aligned}
\label{eq:tilA-v}
\end{align}
One can remove the linear term of $v$ by performing the modular S-transformation
\begin{align}
 \til{A}(z)=-\frac{\ri}{4}\del_{v'}\log\Biggl[
\vartheta_3\Bigl(v'+\frac{\tau'}{2},\tau'\Bigr)^2+
\vartheta_4\Bigl(v'+\frac{\tau'}{2},\tau'\Bigr)^2\Biggr]
\end{align}
with
\begin{align}
 v'=\frac{v}{\tau},\quad\tau'=-\frac{1}{\tau}.
\end{align}
As discussed in \cite{Borot:2009ia},
we could use the S-dual variables $(v',\tau')$ from the beginning.
However, we will not do so and we will continue to use the original variables $(v,\tau)$.

In the following, we will show that
$\om(z)$ in \eqref{eq:om-z}
indeed satisfies the necessary functional relations.

\paragraph{Symmetries of $\boldsymbol{\om(v)}$}
First, let us consider the relation  \eqref{eq:om-inv}.
In terms of $v$,  \eqref{eq:om-inv} is written as
\begin{align}
 \om\Bigl(v+\frac{\tau}{2}\Bigr)=-\om(v).
\label{eq:half-tau}
\end{align}
This is satisfied since $z(v+\tau/2)=1/z(v)$ and
\begin{align}
 \til{A}\Bigl(v+\frac{\tau}{2}\Bigr)=\til{A}(v),
\end{align}
which follows from the identity of Jacobi theta functions
\begin{align}
 \vartheta_4\Bigl(v+\frac{\tau}{2},\tau\Bigr)=\ri e^{-\pi \ri(v+\frac{\tau}{4})}\vartheta_1(v,\tau),\quad
\vartheta_1\Bigl(v+\frac{\tau}{2},\tau\Bigr)=\ri e^{-\pi \ri(v+\frac{\tau}{4})}\vartheta_4(v,\tau).
\end{align}
Then, the $\tau/2$-shift relation
\eqref{eq:half-tau} implies that $\om(v)$ is periodic \eqref{eq:om-c2} with period $\tau$.
One can also show  the relation
\eqref{eq:om-c1}
by using
\begin{align}
\begin{aligned}
 \til{A}\Bigl(-\hf+v\Bigr)&=-\til{A}\Bigl(-\hf-v\Bigr),\\
z\Bigl(-\hf+v\Bigr)&=z\Bigl(-\hf-v\Bigr). 
\end{aligned}
\end{align}

Next let us consider the normalization condition \eqref{eq:om-la}.
So far $\h{\la}$ in \eqref{eq:om-z}
is just a formal parameter in the ansatz of solution.
The relation between $\h{\la}$ and the 't Hooft
coupling 
$\la$ is fixed by the condition \eqref{eq:om-la}. 
Using $\til{A}(0)=\pi/2$,
we find
\begin{align}
 \h{\la}=\la+\frac{1}{8}.
\end{align} 
This shift by $1/8$ is consistent with the 
Fermi gas result \eqref{eq:abc}
\begin{align}
 N-B=N_f\left(\la+\frac{1}{8}\right)-\frac{1}{2N_f},
\end{align}
at the leading order in the 't Hooft limit.

\paragraph{Absence of poles at $\boldsymbol{z=\pm1}$}
We also want to show that the resolvent is regular at $z=\pm 1$.
This requires
\begin{align}
 \til{A}\Bigl(\hf+\frac{\tau}{4}\Bigr)=\pi,\qquad \til{A}\Bigl(-\hf+\frac{\tau}{4}\Bigr)=0.
\label{eq:no-res}
\end{align}
One can show that \eqref{eq:no-res}
is satisfied by using 
\begin{align}
\begin{aligned}
 \del_vz\big|_{z=\pm1}&=2\ri(a^2-1)K,\\
 \del_v\log\vartheta_4\Bigl(\pm\hf+\frac{\tau}{4}\Bigr)&=-\frac{\ri\pi}{2}
\pm \ri (1-a^2)K.
\end{aligned}
\label{eq:Zat1}
\end{align}
The second equality in \eqref{eq:Zat1}
is a consequence of the formula in the
Appendix A of \cite{Grassi:2014vwa}.

\paragraph{Absence of poles at $\boldsymbol{z=\pm \ri}$}
The regularity of resolvent at $z=\pm\ri$ determines $\h{\la}$
as a function of $a$. 
One can show that $A(u)$
in \eqref{eq:tAvsA} is regular at $z=\pm\ri$,
and hence near  $z=\pm\ri$ the resolvent behaves as
\begin{align}
 \lim_{z\to\pm\ri}\om(z)=\frac{1}{2\pi^2}\left(\frac{K'\del_uz}{z(z^2+1)}\right)^2\frac{z^2+1}{z^2-1}+\h{\la}\frac{z^2-1}{z^2+1}+(\text{regular}).
\end{align}
From this behavior, the condition for the absence of pole at $z=\pm\ri$ is
found to be
\begin{align}
 \h{\la}=\frac{t^2}{8\pi^2},
\label{eq:hla-cond}
\end{align}
where $t$ is given by
\begin{align}
 t=K'\del_u z\Big|_{z=\pm\ri}.
\end{align}
Using \eqref{eq:del-z} we find
that $t$ is written as
\begin{align}
 t=(1+a^2)K'.
\label{eq:t-K'}
\end{align}
This reproduces the result in \cite{Grassi:2014vwa}.

\paragraph{Loop equation}
Finally, let us show that $\om(v)$ in \eqref{eq:om-z}
satisfies the loop equation
\eqref{eq:om-loop}.
In \cite{Grassi:2014vwa}, this was shown
implicitly by taking a limit of the resolvent of $O(n)$
matrix model. Here we will show the loop equation
\eqref{eq:om-loop} directly.

To do this, it is convenient to introduce the operator 
$\textbf{T}^{\pm}$  shifting  $v$ by $\pm1$
\begin{align}
 \textbf{T}^\pm\,\om(v)=\om(v\pm1).
\end{align} 
Then the planar loop equation is written as
\begin{align}
 \textbf{L}\,\om(v)=V'(-z),
\label{eq:Lom}
\end{align} 
where $\textbf{L}$ is given by \footnote{For the $O(n)$ model with
$n=-2\cos\pi\nu$,
we have the operator
\begin{align}
 \textbf{L}=\textbf{T}+\textbf{T}^{-1}-n=\textbf{T}^{-1}(\textbf{T}+e^{\pi\ri\nu})
(\textbf{T}+e^{-\pi\ri\nu}).
\end{align}
Then it is natural to decompose 
the resolvent into the eigenfunctions $G_\pm(v)$
with eigenvalues
$\textbf{T}=-e^{\pm\ri\nu}$.
However, for $n=2$
$\textbf{L}$ has a double root at $\textbf{T}=1$ 
which is a somewhat degenerate case.
This is similar to  solving 
the linear differential equation
$(d/dx-1)^2y=0$: it is well known that
one of the solution
$y=xe^x$ is not an eigenfunction of $\frac{d}{dx}$.
One can think of 
the function $\til{A}(z)$ as an analogue
of this solution $y=xe^x$.}
\begin{align}
 \textbf{L}=\textbf{T}+\textbf{T}^{-1}-2.
\end{align}
On the right hand side of \eqref{eq:Lom}, we have also used $\textbf{T}\cdot z=-z$ and $V'(v+1)=V'(-z)$.

For a rational function $f(z)$ of $z$,
the action of $\textbf{L}$ reads
\begin{align}
\textbf{L}\, f(z)=2f(-z)-2f(z).
\end{align}
However, we should be careful about
$\til{A}(z)$ which transforms inhomogeneously
under $\textbf{T}^\pm$
\begin{align}
 \textbf{T}^{\pm}\,\til{A}(z)=\til{A}(z)\pm \pi,
\end{align}
due to the linear term $\pi  v$ in \eqref{eq:tilA-v}.
Then we find
\begin{align}
 \textbf{L}\,\til{A}(z)^2=\big(\til{A}(z)+\pi\big)^2+\big(\til{A}(z)-\pi\big)^2-2\til{A}(z)^2
=2\pi^2.
\label{eq:LonA^2}
\end{align} 

Now we are ready to prove \eqref{eq:om-loop}.
It is natural to decompose $\om(z)$ in \eqref{eq:om-z} into 
three parts
\begin{align}
 \om(z)=\om_1(z)+\om_2(z)+\om_3(z),
\end{align}
where
\begin{align}
 \begin{aligned}
  \om_1(z)&=\frac{\til{A}(z)^2}{2\pi^2}\frac{z^2+1}{z^2-1}
=\frac{\til{A}(z)^2}{4\pi^2}\Bigl[V'(z)+V'(-z)\Bigr],\\
\om_2(z)&=-\qu V'(-z),\\
\om_3(z)&=\h{\la}\,\frac{z^2-1}{z^2+1}.
 \end{aligned}
\end{align}
One can easily show that the action of $\textbf{L}$ on $\om_{1,2,3}$ is given by
\begin{align}
  \begin{aligned}
   \textbf{L}\,\om_1(z)&=\hf\Bigl[V'(z)+V'(-z)\Bigr],\\
\textbf{L}\,\om_2(z)&=-\hf \Bigl[V'(z)-V'(-z)\Bigr],\\
 \textbf{L}\,\om_3(z)&=0.
  \end{aligned}
\label{eq:Lon-om}
\end{align}
where we have used \eqref{eq:LonA^2}
in the first line.
Adding these three equations \eqref{eq:Lon-om}, 
finally we arrive at the desired loop equation \eqref{eq:Lom}.

\subsection{Planar free energy}
In \cite{Grassi:2014vwa}, the second derivative of 
the planar free energy is 
found explicitly as
\begin{align}
 \frac{\del^2 F_0}{\del\la^2}=-\frac{2\pi K}{K'}=-\frac{2\pi\ri}{\tau}.
\label{eq:F0d2}
\end{align}
From \eqref{eq:hla-cond}, \eqref{eq:t-K'} and \eqref{eq:F0d2}, we find
that the derivative of $\la$ and $\frac{\del F_0}{\del\la}$
with respect to $t$ have a simple form
\begin{align}
 \begin{aligned}
  \frac{\del \la}{\del t}&=\frac{t}{4\pi^2}=\frac{(1+a^2)K'}{4\pi^2},\\
\frac{\del^2 F_0}{\del t\del\la}&=-\frac{(1+a^2)K}{2\pi}.
 \end{aligned}
\label{eq:AB-peri}
\end{align}
Note that the role of A-period and B-period
is opposite from the standard definition.
Using this relation \eqref{eq:AB-peri}, one can
find the planar free energy $F_0$ as a function of 't Hooft coupling $\la$.
In particular, in the small $\la$ or large $\la$ regime
$F_0$ can be explicitly found as a power series.

Let us first consider the large $\la$ behavior of $F_0$.
In the large $\la$ limit the size of the cut $[-T,T]$ in the original variable
$x$ becomes large,
which implies $a=e^{-T}\to0$. 
More precisely, we find that the large $t$
expansion of $a$ is given by
\begin{align}
 a=2e^{-\frac{t}{2}}\Biggl[1+2te^{-t}+2(5t^2-3t-1)e^{-2t}
+\frac{4t}{3}(49t^2-63t-3)e^{-3t}+\mathcal{O}(e^{-4t})\Biggr].
\label{eq:a-larget}
\end{align}
Note that $t$ and the shifted 't Hooft coupling $\h{\la}=\la+1/8$ 
are related by \eqref{eq:hla-cond}
and the exponential correction $e^{-t}$ in \eqref{eq:a-larget} is 
identified with the worldsheet instanton factor \eqref{eq:order-inst}
\begin{align}
 e^{-t}=e^{-2\pi\rt{2\h{\la}}}.
\end{align}
Then, integrating the relation \eqref{eq:AB-peri}
the planar free energy becomes
\begin{align}
 -F_0=\frac{t^3}{48 \pi ^2}-a_0+\frac{t+1}{4 \pi ^2}e^{-t}+\frac{8 t^2+14
   t+7}{32 \pi ^2}e^{-2t}
+\frac{18 t^3+27 t^2+21 t+7}{27 \pi ^2}
e^{-3t}+\mathcal{O}(e^{-4t}),
\end{align}
where $a_0$ is a constant coming from
the constant term $A$ in the perturbative part \eqref{eq:Jpert}
of grand potential \cite{HO}
\begin{align}
 a_0=\lim_{N_f\to\infty}\frac{A}{N_f^2}=-\frac{\zeta(3)}{8\pi^2}+\frac{1}{8}\log2.
\end{align}
On the other hand, in the small $\la$ limit 
the size of cut $[-T,T]$ in the $x$-variable becomes small, which implies $a\to1$.
From \eqref{eq:hla-cond} and \eqref{eq:t-K'},
we find that the small $\la$ expansion of $\log a$ is given by
\begin{align}
 \log a=-4\rt{\la}\left(1-\frac{\lambda }{6}
+\frac{43 \lambda ^2}{40}
-\frac{621 \lambda
   ^3}{112}+\frac{35027 \lambda ^4}{1152}+\mathcal{O}(\la^5)\right),
\label{eq:loga-exp}
\end{align}
and the free energy becomes
\begin{align}
-F_0= -\frac{\lambda ^2}{2}  \log \la+\frac{3 \lambda
   ^2}{4}+\frac{\lambda ^3}{2}-\frac{19
   \lambda ^4}{24}+\frac{9 \lambda ^5}{4}
-\frac{991 \lambda ^6}{120}+\mathcal{O}(\la^7).
\end{align}

\section{'t Hooft limit of Wilson loops}\label{sec:tHooft}
In this section, we consider the 't Hooft expansion of  {\it normalized} Wilson loop VEV
\begin{align}
 \frac{W_m(N,N_f)}{Z(N,N_f)}=\sum_{g=0}^\infty N_f^{1-2g}
W_m^{(g)}.
\label{eq:genus-Wm}
\end{align}
Note that our normalization of VEV is different from \cite{Grassi:2014vwa}
(see footnote \ref{foot:normalization} for our definition).
As we will see below, we find a perfect agreement between the matrix model result
and the Fermi gas result for the genus-zero
part $W_m^{(g=0)}$ of Wilson loop in \eqref{eq:genus-Wm}.

\subsection{Results of matrix model}
We can read off the genus-zero VEV of Wilson loops
from the small $z$ expansion of the resolvent \eqref{eq:om-z}\footnote{Using the symmetry
$\om(z^{-1})=-\om(z)$, one can read off the Wilson loop VEVs from 
the large $z$ expansion of resolvent as well
\begin{align}
 \om(z)=\la+2\sum_{m=1}^\infty \frac{W_m^{(g=0)}}{z^{m}}.
\end{align}
}
\begin{align}
 \om(z)=-\la-2\sum_{m=1}^\infty W_m^{(g=0)}z^m.
\label{eq:om-smallz}
\end{align}
To write down the small $z$ expansion of $\om(z)$ in \eqref{eq:om-z}, let us first consider the small 
$u$ expansion
of $\til{A}(z)$ in \eqref{eq:tAvsA}
\begin{align}
 \til{A}(z)=\frac{\pi}{2}+Iu+t\Bigl[
-\frac{2}{3} a^2 \left(a^2+1\right) u^3
+\frac{2}{15} a^2 \left(a^2+1\right) \left(a^4+6
   a^2+1\right) u^5
+\mathcal{O}(u^7)\Bigr],
\label{eq:tilA-exp}
\end{align}
where $t$  is defined in \eqref{eq:t-K'}
and the factor $I$
is given by
\begin{align}
 I=\frac{\pi}{2K}+t\left(1-\frac{E}{(1+a^2)K}\right).
\label{eq:I-def}
\end{align}
Here $E=E(k^2)$ denotes the elliptic integral of the second kind.
By inverting the relation $z=a\,\text{sn}(u)$ in \eqref{eq:z-snu}, we can  write down
the small $z$ expansion of $u$
\begin{align}
 u=\frac{z}{a}+\frac{\left(a^4+1\right) z^3}{6 a^3}+\frac{\left(3 a^8+2
   a^4+3\right) z^5}{40 a^5}+\mathcal{O}(z^7).
\label{eq:u-in-z}
\end{align}
Combining \eqref{eq:tilA-exp}
and \eqref{eq:u-in-z},
we find that $\til{A}(z)$
is expanded as
\begin{align}
 \begin{aligned}
  \til{A}(z)&=\frac{\pi}{2}+\frac{I}{a}z+\Biggl[\frac{\left(a^4+1\right) I}{6 a^3}-\frac{2
   \left(a^2+1\right) t}{3 a}\Biggr]z^3\\
&+\Biggl[\frac{\left(3 a^8+2 a^4+3\right) I}{40
   a^5}-\frac{\left(a^2+1\right) \left(a^4-4 a^2+1\right)
   t}{5 a^3}\Biggr]z^5+\mathcal{O}(z^7).
 \end{aligned}
\label{eq:tilA-in-z}
\end{align}
We notice that
the coefficients in this expansion \eqref{eq:tilA-in-z}
are some linear combinations of $I$ and $t$.
Plugging this expansion \eqref{eq:tilA-in-z}
into \eqref{eq:om-z} 
and read off the coefficient of $z^m$ in \eqref{eq:om-smallz},
we can find the planar VEV of winding Wilson loops $W_m^{(g=0)}$
up to arbitrary winding number $m$, in principle.
For instance, the planar VEV of  Wilson loop
in the fundamental representation is given by
\begin{align}
 W_{\tableau{1}}^{(g=0)}
=-\frac{1}{4}+\frac{I}{4\pi a}.
\label{eq:fund-gzero} 
\end{align}
This agrees with the result of \cite{Grassi:2014vwa}.
For the higher winding numbers we find
\begin{align}
\begin{aligned}
 W_2^{(g=0)}&=-\la-\qu
+\frac{I^2}{4\pi^2a^2},
\\
W_3^{(g=0)}&=-\qu-\frac{t(1+a^2)}{6\pi a}
+\frac{(1+12a^2+a^4)I}{24\pi a^3},\\
W_4^{(g=0)}&=\lambda-\frac{\left(a^2+1\right) tI}{3 \pi ^2
   a^2}+\frac{\left(a^4+6 a^2+1\right) I^2}{12 \pi ^2
   a^4},\\
W_5^{(g=0)}&=-\qu-\frac{(a^2+1)\left(3 a^4+8a^2+3\right) t}{60 \pi 
   a^3}
+\frac{\left(9 a^8+40 a^6+246 a^4+40 a^2+9\right) I}{480 \pi 
   a^5},\\
W_6^{(g=0)}&=-\la-\qu
+\frac{\left(a^2+1\right)^2 t^2}{9 \pi ^2
   a^2}-\frac{\left(a^2+1\right) \left(7 a^4+12 a^2+7\right)
   tI}{45 \pi ^2 a^4}\\
&\quad+\frac{\left(8 a^8+30 a^6+97 a^4+30
   a^2+8\right) I^2}{180 \pi ^2 a^6}.
\end{aligned}
\label{eq:Wg=0}
\end{align}
We observe that the planar VEV of winding Wilson loop is a linear polynomial of $I$
for odd $m$, and quadratic in $I$ for even $m$. This structure
originates from the linear dependence of $\til{A}(z)$ 
on $I$ in \eqref{eq:tilA-in-z}. 

\paragraph{Small $\boldsymbol{\lambda}$ expansion}
From the small $\la$ expansion of $\log a$ in \eqref{eq:loga-exp},
one can easily find the small $\la$ expansion of Wilson loop VEVs in \eqref{eq:Wg=0}.
For general winding number $m$,
we find that the Wilson loop VEVs are expanded as
\begin{align}
\begin{aligned}
 W_m^{(g=0)}&=\la+2m^2\la^2+\frac{4}{3}m^2(m^2-1)\la^3
+\frac{2}{9}m^2(2m^4-7m^2+23)\la^4\\
&+\frac{4}{15} m^2(m-2) \left(7 m^4-47 m^3+151 m^2-213
   m+162\right)\la^5+\mathcal{O}(\la^6).
\end{aligned}
\label{eq:Wm-small-la}
\end{align}
For instance, the small $\la$ expansion of the fundamental representation is
given by
\begin{align}
  W_{\tableau{1}}^{(g=0)}&=\lambda+2\la^2
+4\la^4-16\la^5+\mathcal{O}(\la^6),
\end{align}
which reproduces the result in \cite{Grassi:2014vwa}\footnote{Note that our normalization
is different from \cite{Grassi:2014vwa} by a factor of $\la$.}.
We have checked that \eqref{eq:Wm-small-la}
correctly reproduces the expansion of $W_m^{(g=0)}$
for $m=1,\cdots,6$ in \eqref{eq:Wg=0}. We conjecture that \eqref{eq:Wm-small-la}
holds for any winding number $m$.
It would be interesting to reproduce this expansion \eqref{eq:Wm-small-la} from the 
perturbative calculation of matrix model along the lines of
\cite{Grassi:2014vwa}.

\paragraph{Large $\boldsymbol{t}$ expansion}
One can also study the 
large 't Hooft coupling, or large $t$ behavior of Wilson loop VEVs \eqref{eq:Wg=0}.
This large $t$ regime is directly related to the Fermi gas result, which we will
consider in the next subsection.

The large $t$ behavior of $W_m^{(g=0)}$ can be found from the small $a$
expansion of $I$ in \eqref{eq:I-def}
\begin{align}
 \begin{aligned}
  I&=1+ta^2-\frac{1+2t}{4}a^4+\frac{t}{2}a^6
-\frac{5+28t}{64}a^8+\mathcal{O}(a^{10}),
 \end{aligned}
\end{align}
together with the large $t$ expansion of $a$ in \eqref{eq:a-larget}.
For the first three 
winding numbers $m=1,2,3$, the large $t$ expansion of
$W_m^{(g=0)}$ is given by
\begin{align}
\begin{aligned}
 W_{\tableau{1}}^{(g=0)}
&=-\qu+\frac{e^{\hf t}}{8\pi}\Biggl[1+2te^{-t}+
   2\left(t^2-t-1\right)e^{-2t}+\frac{4t(5t^2-9t-3)}{3}e^{-3t}+\mathcal{O}(e^{-4t})\Biggr],\\
 W_2^{(g=0)}&=-\la-\qu
+\frac{e^t}{16\pi^2}\Biggl[1+4te^{-t}+4(2t^2-t-1)e^{-2t}+\mathcal{O}(e^{-3t})\Biggr], \\
 W_3^{(g=0)}&=-\qu+\frac{e^{\frac{3t}{2}}}{192\pi}
\Biggl[1-6(3t-8)e^{-t}+6(3t^2 - 17 t + 3)e^{-2t}+\mathcal{O}(e^{-3t})\Biggr].
\end{aligned}
\label{eq:Wplanar-t}
\end{align}

\subsection{Comparison with Fermi gas}
Let us compare the matrix model results \eqref{eq:Wplanar-t}
with the Fermi gas results in section \ref{sec:inst}.
In the grand canonical picture, the 't Hooft limit
is given by
\begin{align}
 N_f,\mu\to\infty,\quad \frac{\mu}{N_f}:\text{fixed}
\label{eq:thooft-mu}
\end{align}
As discussed in \cite{Grassi:2014vwa},
at the level of genus-zero the Wilson loop VEV in the canonical picture
can be obtained by plugging the saddle point value $\mu_*$
of the chemical potential
into the grand canonical VEV 
$W_m(\mu_*,N_f)$ of Wilson loop.
However, to study the instanton corrections 
we have to include the exponentially small corrections to the
saddle point value  $\mu_*$ of the chemical potential,
beyond the perturbative expression in \eqref{eq:mu-saddle}.
This is achieved by identifying the saddle point value
$\mu_*$ with the derivative of the planar free energy $F_0$ \cite{Grassi:2014vwa}
\begin{align}
 \frac{\mu_*}{N_f}= -\frac{\del F_0}{\del\la}
=\frac{t}{4}-e^{-t}-\hf(4t+3)e^{-2t}
-\frac{4}{3}  \left(6 t^2+3 t+1\right)e^{-3 t}
+\mathcal{O}(e^{-4t}).
\label{eq:mu*-inst}
\end{align} 
Plugging this expansion of $\mu_*$ \eqref{eq:mu*-inst}
into the grand canonical VEV $\h{W}_{m}(\mu=\mu_*,N_f)$
in section \ref{sec:inst}
(eqs.\eqref{eq:fund-inst-fermi}, \eqref{eq:w2-inst},
 \eqref{eq:w3-inst} for $m=1,2,3$, respectively), we have confirmed
that the Fermi gas results perfectly match
the matrix model results \eqref{eq:Wplanar-t} in the planar limit.

Let us take a closer look at the correspondence between the
Fermi gas results and the matrix model results.
The perturbative part \eqref{eq:Wm-scale}
in the Fermi gas picture corresponds to the term $I^\ga/a^m$
in the matrix model result
\eqref{eq:Wg=0}, where
$\ga=1,2$ for odd $m$ and even $m$, respectively.
We find that the coefficient of this term in $W_m^{(g=0)}$
is given by
\begin{align}
 \frac{2^m\Ga\big(\frac{m}{2}\big)^2}{8\pi^2 \Ga(m+1)}\frac{I^\ga}{a^m}.
\label{eq:Wm-lead-I}
\end{align}
In the large $t$ limit this term
\eqref{eq:Wm-lead-I} becomes
\begin{align}
 \frac{\Ga\big(\frac{m}{2}\big)^2}{8\pi^2 \Ga(m+1)}e^{\frac{mt}{2}}.
\label{eq:Wm-lead}
\end{align}
One can see that this matrix model result \eqref{eq:Wm-lead}
is correctly reproduced
from the perturbative part
in the Fermi gas picture \eqref{eq:Wm-scale}
in the 't Hooft limit
\eqref{eq:thooft-mu}
\begin{align}
\lim_{N_f\to\infty}\frac{1}{N_f}c_m(N_f)e^{\frac{2m\mu}{N_f}}
=\frac{\Ga\big(\frac{m}{2}\big)^2}{8\pi^2 \Ga(m+1)}e^{\frac{2m\mu}{N_f}},
\end{align}
where $\mu$ should be identified with
the saddle point value $\mu_*$ in \eqref{eq:mu*-inst}.

Also, we observe that the matrix model result $W_m^{(g=0)}$
\eqref{eq:Wplanar-t} contains several pieces with different scalings
in the large $t$ limit, 
which naturally corresponds to the similar decomposition in the Fermi gas picture. 
For instance,
the constant $-1/4$ in the fundamental representation
\eqref{eq:fund-gzero} corresponds to the first term in the 
decomposition \eqref{eq:fund-decomp} observed in the Fermi gas picture.
Similarly, the first two terms in the $m=2$ VEV in \eqref{eq:Wplanar-t}
corresponds to the genus-zero part of the first term in
\eqref{eq:decom-w2}
\begin{align}
 -N+2B=N_f\Bigl(-\la-\qu\Bigr)+\frac{1}{N_f}.
\end{align}
For $m=3$, the constant term $-1/4$ in 
$W_3^{(g=0)} $\eqref{eq:fund-gzero}
agrees with the Fermi gas result \eqref{eq:decom-w3}, 
which further suggests
the following decomposition of planar VEV
\begin{align}
 W_3^{(g=0)}=-\qu+3\til{W}_1^{(g=0)}+\til{W}_3^{(g=0)}
\end{align}
with
\begin{align}
 \til{W}_3^{(g=0)}=\frac{e^{\frac{3t}{2}}}{192\pi}\Biggl[1-6(3t+4)e^{-t}+6(3t^2 - 17 t + 3)e^{-2t}
+\mathcal{O}(e^{-3t})\Biggr].
\end{align}

We should stress that our Fermi gas
results in section \ref{sec:inst}
have all-order information of the genus expansion.
In other words, one can predict the higher genus amplitudes
$W_m^{(g)}$ from the Fermi gas results.
For instance, from \eqref{eq:fund-inst-fermi} the genus-one amplitude 
of the fundamental representation is given by
\begin{align}
\begin{aligned}
 W_{\tableau{1}}^{(g=1)}&=4\pi e^{\hf t}\Biggl[\frac{1}{8 t^2}-\frac{3}{32 t}+\frac{1}{48}
+\left(\frac{3}{8 t}-\frac{7}{48}\right)e^{-t}+
\left(-\frac{1}{4 t^2}-\frac{39}{16
   t}+\frac{49}{16}-\frac{31 t}{48}-\frac{t^2}{24}\right)e^{-2t}\\
&\qquad+\left(\frac{29}{4 t}-\frac{287}{8}+\frac{185
   t}{8}-\frac{29 t^2}{24}-\frac{17 t^3}{18}\right)e^{-3t}+\mathcal{O}(e^{-4t})\Biggr]. 
\end{aligned}
\end{align}
It would be interesting to see if  this is reproduced from the matrix model
calculation at genus-one.

\section{Conclusions}\label{sec:conclude}
In this paper, we have studied the Wilson loops in the $N_f$
matrix model from the Fermi gas approach.
We have determined the first few worldsheet
instanton corrections to the winding Wilson loops 
for the winding number $m=1,2,3$,
and found that our Fermi gas result is consistent with
the planar limit of matrix model result.
We find that the Wilson loop VEVs can be decomposed
into several pieces with different scaling behavior in the large $N$
limit.
Also, we conjecture that the 
grand canonical VEVs of winding Wilson loops do not  
receive ``pure'' membrane instanton corrections
except for the bound state contributions.
This is reminiscent of the instanton corrections to
the 1/2 BPS Wilson loops in the ABJM theory \cite{HHMO,Hatsuda:2016rmv}.

There are many interesting open problems.
To study the partition functions and Wilson loops in the $N_f$
matrix model further, it is very important to
understand the structure of bound states.
In the case of ABJM theory, the effect of bound states can be removed
by introducing the ``effective'' chemical potential
$\mu_{\text{eff}}$ \cite{Hatsuda:2013gj}, which in turn is related to the
quantum period of the quantized mirror curve of local
$\mathbb{P}^1\times \mathbb{P}^1$ \cite{Aganagic:2011mi,HMMO}.
It would be interesting to see if one can define a similar
``effective'' chemical potential in the $N_f$ matrix model as well.

Our study was limited to the 
single trace winding Wilson loops.
It would be important to develop a technique to
analyze the Wilson loops in general representations
and study their instanton corrections.
In particular, it would be interesting to 
consider the Wilson loops in representations with large dimensions,
which are expected to be holographically dual to
certain configurations of D-branes.
Also, it would be interesting to study 
implications of our findings to the
mirror symmetry between Wilson loops and vortex loops 
in 3d $\mathcal{N}=4$ theories \cite{Assel:2015oxa}.

\vskip10mm
\acknowledgments
I would like to thank Alba Grassi, Yasuyuki Hatsuda,
and Marcos Mari\~no for correspondence and discussion.
This work was supported in part by JSPS KAKENHI Grant Number
16K05316, and JSPS Japan-Hungary and Japan-Russia bilateral
joint research projects.

%%%%%%%%%%%%%%%%%%%%%%%
\appendix

\section{Exact values of Wilson loop VEVs}\label{app:exact}
In this appendix, we list some exact values of Wilson loop VEVs for winding number $m=1,2,3$.
\subsection{Fundamental representation}
Below we list the exact values of $W_{\tableau{1}}(N,N_f)$.

For $N_f=4$ we find
\begin{align}
 \begin{aligned}
  W_{\tableau{1}}(1,4)&=\frac{1}{12 \pi },\\
W_{\tableau{1}}(2,4)&=\frac{32-3 \pi ^2}{1536 \pi ^2},\\
W_{\tableau{1}}(3,4)&=\frac{32 \pi ^2-315}{161280 \pi ^3},\\
W_{\tableau{1}}(4,4)&=\frac{-16800+11029 \pi ^2-945 \pi ^4}{61931520 \pi ^4}.
 \end{aligned}
\end{align}

For $N_f=6$ we find
\begin{align}
 \begin{aligned}
  W_{\tableau{1}}(1,6)&=\frac{1}{80 \pi },\\
W_{\tableau{1}}(2,6)&=\frac{448-45 \pi ^2}{122880 \pi ^2},\\
W_{\tableau{1}}(3,6)&=\frac{2283 \pi ^2-22528}{389283840 \pi ^3},\\
W_{\tableau{1}}(4,6)&=\frac{-93040640+50977776 \pi ^2-4209975 \pi
   ^4}{5231974809600 \pi ^4}.
 \end{aligned}
\end{align}

For $N_f=8$ we find
\begin{align}
 \begin{aligned}
  W_{\tableau{1}}(1,8)&=\frac{1}{420 \pi },\\
W_{\tableau{1}}(2,8)&=\frac{77824-7875 \pi ^2}{103219200 \pi ^2},\\
W_{\tableau{1}}(3,8)&=\frac{7419 \pi ^2-73216}{141699317760 \pi ^3},\\
W_{\tableau{1}}(4,8)&=\frac{-21615968518144+10372082726400 \pi ^2-829002549375 \pi
   ^4}{18282612774666240000 \pi ^4}.
 \end{aligned}
\end{align}
\subsection{Winding number $m=2$}
Here we list the exact values of $W_{2}(N,N_f)$.

For $N_f=6$ we find
\begin{align}
 \begin{aligned}
  W_2(1,6)&=\frac{1}{20 \pi },\\
 W_2(2,6)&=-\frac{7 \left(45 \pi ^2-448\right)}{122880 \pi ^2},\\
 W_2(3,6)&=\frac{1623 \pi ^2-16016}{18923520 \pi ^3},\\
W_2(4,6)&=\frac{-22392832+18581400 \pi ^2-1652805 \pi ^4}{174399160320
   \pi ^4}.
 \end{aligned}
\end{align}

For $N_f=8$ we find
\begin{align}
 \begin{aligned}
  W_2(1,8)&=\frac{1}{168 \pi },\\
W_2(2,8)&=\frac{699904-70875 \pi ^2}{103219200 \pi ^2},\\
W_2(3,8)&=\frac{812775 \pi ^2-8021728}{147603456000 \pi ^3},\\
W_2(4,8)&=\frac{-97971039502336+70939141382400 \pi ^2-6181868818125
   \pi ^4}{9141306387333120000 \pi ^4}.
 \end{aligned}
\end{align}

\subsection{Winding number $m=3$}
This is the list of the exact values of $W_{3}(N,N_f=8)$
\begin{align}
 \begin{aligned}
  W_3(1,8)&=\frac{1}{28 \pi },\\
 W_3(2,8)&=\frac{6075 \pi ^2-59392}{14745600 \pi ^2},\\
W_3(3,8)&=\frac{2760375 \pi ^2-27243008}{193226342400 \pi ^3},\\
W_3(4,8)&=\frac{15900762701824-5780821056000 \pi ^2+422482685625 \pi
   ^4}{2031401419407360000 \pi ^4}.
 \end{aligned}
\end{align}

\section{A curious observation for $N_f=4$}\label{app:curious}
We find a curious relation between 
 the VEV of fundamental Wilson loop
$\til{W}_{\tableau{1}}(N,N_f)$ in \eqref{eq:fund-decomp}
with $N_f=4$ and
the partition function $Z_{p,q}(N,k)$
of a certain circular quiver Chern-Simons-matter theory 
with the gauge $U(N)_{k}\times U(N)_0^{p-1}\times
U(N)_{-k}\times U(N)^{q-1}_0$,
where the subscripts denote the Chern-Simons level.
We find that  $Z_{p,q}(N,k)$ with 
$(p,q,k)=(1,2,2)$ 
is exactly equal to $\til{W}_{\tableau{1}}(N,N_f=4)$.
For instance, the first three terms are
\begin{align}
 \begin{aligned}
  \til{W}_{\tableau{1}}(1,4)=\frac{1}{8 \pi },\quad
\til{W}_{\tableau{1}}(2,4)=\frac{\pi ^2-8}{1024 \pi ^2},\quad
\til{W}_{\tableau{1}}(3,4)=\frac{61 \pi ^2-600}{368640 \pi ^3},
 \end{aligned}
\end{align}
which agree with $Z_{1,2}(N,2)$ computed in \cite{Moriyama:2014gxa,Hatsuda:2015lpa}.
Furthermore, by looking at the non-perturbative part of grand potential
of $N_f=4$ in \cite{HO}
and $(p,q,k)=(1,2,2)$ model in \cite{Moriyama:2014gxa,Hatsuda:2015lpa}
\begin{align}
\begin{aligned}
 J(\mu,N_f=4)&=-\frac{2(\mu+1)}{\pi}e^{-\mu}+\left[-\frac{10\mu^2+7\mu+7/2}{\pi^2}+1\right]e^{-2\mu}
-\frac{88\mu+52/3}{3\pi}e^{-3\mu} ,\\
J_{1,2}(\mu,k=2)&=\frac{2(\mu+1)}{\pi}e^{-\mu}+\left[-\frac{10\mu^2+7\mu+7/2}{\pi^2}+1\right]e^{-2\mu}
+\frac{88\mu+52/3}{3\pi}e^{-3\mu},
\end{aligned}
\end{align}
we find a curious similarity with the 
grand potential of local $\mathbb{F}_2$ with mass parameter $m_{\mathbb{F}_2}=0$
in the maximal supersymmetric case $\hbar=2\pi$ \cite{Gu:2015pda}
\begin{align}
 J_{\mathbb{F}_2}(\mu,m_{\mathbb{F}_2}=0)=-\frac{2\mu+1}{2\pi}e^{-2\mu}
+\left[-\frac{10\mu^2+7/2\mu+7/8}{\pi^2}+\frac{7}{4}\right]e^{-4\mu}
-\frac{44\mu+13/3}{3\pi}e^{-6\mu}.
\end{align}
We observe that
\begin{align}
 J(\mu,N_f=4)\sim 4J_{\mathbb{F}_2}(\mu/2,m_{\mathbb{F}_2}=0),
\end{align}
except for the difference of the coefficient of $e^{-2\mu}$ without the $1/\pi^2$ factor:
it is $1$ for the $N_f=4$ model while $7/4$ for the local $\mathbb{F}_2$.
It would be interesting to
see if there is a connection between the
$(p,q)=(1,2)$ model and the topological string on local $\mathbb{F}_2$.

%%%%%%%%%%%%%%%%%%%%%%%%


\begin{thebibliography}{99}
\bibitem{MP} 
  M.~Marino and P.~Putrov,
  ``ABJM theory as a Fermi gas,''
  J.\ Stat.\ Mech.\  {\bf 1203}, P03001 (2012)
  doi:10.1088/1742-5468/2012/03/P03001
  [arXiv:1110.4066 [hep-th]].
  %%CITATION = doi:10.1088/1742-5468/2012/03/P03001;%%
  %130 citations counted in INSPIRE as of 09 Jul 2016

%\cite{Aharony:2008ug}
\bibitem{Aharony:2008ug} 
  O.~Aharony, O.~Bergman, D.~L.~Jafferis and J.~Maldacena,
  ``N=6 superconformal Chern-Simons-matter theories, M2-branes and their gravity duals,''
  JHEP {\bf 0810}, 091 (2008)
  doi:10.1088/1126-6708/2008/10/091
  [arXiv:0806.1218 [hep-th]].
  %%CITATION = doi:10.1088/1126-6708/2008/10/091;%%


%\cite{Aharony:2008gk}
\bibitem{Aharony:2008gk} 
  O.~Aharony, O.~Bergman and D.~L.~Jafferis,
  ``Fractional M2-branes,''
  JHEP {\bf 0811}, 043 (2008)
  doi:10.1088/1126-6708/2008/11/043
  [arXiv:0807.4924 [hep-th]].
  %%CITATION = doi:10.1088/1126-6708/2008/11/043;%%

%\cite{Drukker:2010nc}
\bibitem{Drukker:2010nc} 
  N.~Drukker, M.~Marino and P.~Putrov,
  ``From weak to strong coupling in ABJM theory,''
  Commun.\ Math.\ Phys.\  {\bf 306}, 511 (2011)
  doi:10.1007/s00220-011-1253-6
  [arXiv:1007.3837 [hep-th]].
  %%CITATION = doi:10.1007/s00220-011-1253-6;%%

%\cite{Hatsuda:2015gca}
\bibitem{Hatsuda:2015gca} 
  Y.~Hatsuda, S.~Moriyama and K.~Okuyama,
  ``Exact instanton expansion of the ABJM partition function,''
  PTEP {\bf 2015}, no. 11, 11B104 (2015)
  doi:10.1093/ptep/ptv145
  [arXiv:1507.01678 [hep-th]].
  %%CITATION = doi:10.1093/ptep/ptv145;%%

%\cite{Marino:2016new}
\bibitem{Marino:2016new} 
  M.~Marino,
  ``Localization at large N in Chern-Simons-matter theories,''
  arXiv:1608.02959 [hep-th].
  %%CITATION = ARXIV:1608.02959;%%

\bibitem{KMSS} 
  A.~Klemm, M.~Marino, M.~Schiereck and M.~Soroush,
  ``ABJM Wilson loops in the Fermi gas approach,''
  Z.\ Naturforsch.\ A {\bf 68}, 178 (2013)
  %doi:10.5560/ZNA.2012-0118
  [arXiv:1207.0611 [hep-th]].
  %%CITATION = doi:10.5560/ZNA.2012-0118;%%

\bibitem{HHMO} 
  Y.~Hatsuda, M.~Honda, S.~Moriyama and K.~Okuyama,
  ``ABJM Wilson Loops in Arbitrary Representations,''
  JHEP {\bf 1310}, 168 (2013)
  doi:10.1007/JHEP10(2013)168
  [arXiv:1306.4297 [hep-th]].
  %%CITATION = doi:10.1007/JHEP10(2013)168;%%
  %28 citations counted in INSPIRE as of 27 Jun 2016

%\cite{Matsumoto:2013nya}
\bibitem{Matsumoto:2013nya} 
  S.~Matsumoto and S.~Moriyama,
  ``ABJ Fractional Brane from ABJM Wilson Loop,''
  JHEP {\bf 1403}, 079 (2014)
  doi:10.1007/JHEP03(2014)079
  [arXiv:1310.8051 [hep-th]].
  %%CITATION = doi:10.1007/JHEP03(2014)079;%%

%\cite{Hatsuda:2016rmv}
\bibitem{Hatsuda:2016rmv} 
  Y.~Hatsuda and K.~Okuyama,
  ``Exact results for ABJ Wilson loops and open-closed duality,''
  arXiv:1603.06579 [hep-th].
  %%CITATION = ARXIV:1603.06579;%%

%\cite{Matsuno:2016jjp}
\bibitem{Matsuno:2016jjp} 
  S.~Matsuno and S.~Moriyama,
  ``Giambelli Identity in Super Chern-Simons Matrix Model,''
  arXiv:1603.04124 [hep-th].
  %%CITATION = ARXIV:1603.04124;%%



%\cite{Kiyoshige:2016lno}
\bibitem{Kiyoshige:2016lno} 
  K.~Kiyoshige and S.~Moriyama,
  ``Dualities in ABJM Matrix Model from Closed String Viewpoint,''
  arXiv:1607.06414 [hep-th].


%\cite{Mezei:2013gqa}
\bibitem{Mezei:2013gqa} 
  M.~Mezei and S.~S.~Pufu,
  ``Three-sphere free energy for classical gauge groups,''
  JHEP {\bf 1402}, 037 (2014)
  doi:10.1007/JHEP02(2014)037
  [arXiv:1312.0920 [hep-th]].
  %%CITATION = doi:10.1007/JHEP02(2014)037;%%

%\cite{Grassi:2014vwa}
\bibitem{Grassi:2014vwa} 
  A.~Grassi and M.~Marino,
  ``M-theoretic matrix models,''
  JHEP {\bf 1502}, 115 (2015)
  doi:10.1007/JHEP02(2015)115
  [arXiv:1403.4276 [hep-th]].
  %%CITATION = doi:10.1007/JHEP02(2015)115;%%

\bibitem{HO} 
  Y.~Hatsuda and K.~Okuyama,
  ``Probing non-perturbative effects in M-theory,''
  JHEP {\bf 1410}, 158 (2014)
  doi:10.1007/JHEP10(2014)158
  [arXiv:1407.3786 [hep-th]].
  %%CITATION = doi:10.1007/JHEP10(2014)158;%%


%\cite{Intriligator:1996ex}
\bibitem{Intriligator:1996ex} 
  K.~A.~Intriligator and N.~Seiberg,
  ``Mirror symmetry in three-dimensional gauge theories,''
  Phys.\ Lett.\ B {\bf 387}, 513 (1996)
  doi:10.1016/0370-2693(96)01088-X
  [hep-th/9607207].
  %%CITATION = doi:10.1016/0370-2693(96)01088-X;%%

%\cite{deBoer:1996mp}
\bibitem{deBoer:1996mp} 
  J.~de Boer, K.~Hori, H.~Ooguri and Y.~Oz,
  ``Mirror symmetry in three-dimensional gauge theories, quivers and D-branes,''
  Nucl.\ Phys.\ B {\bf 493}, 101 (1997)
  doi:10.1016/S0550-3213(97)00125-9
  [hep-th/9611063].
  %%CITATION = doi:10.1016/S0550-3213(97)00125-9;%%


%\cite{Hatsuda:2012dt}
\bibitem{Hatsuda:2012dt} 
  Y.~Hatsuda, S.~Moriyama and K.~Okuyama,
  ``Instanton Effects in ABJM Theory from Fermi Gas Approach,''
  JHEP {\bf 1301}, 158 (2013)
  doi:10.1007/JHEP01(2013)158
  [arXiv:1211.1251 [hep-th]].
  %%CITATION = doi:10.1007/JHEP01(2013)158;%%


%\cite{Kapustin:2009kz}
\bibitem{Kapustin:2009kz} 
  A.~Kapustin, B.~Willett and I.~Yaakov,
  ``Exact Results for Wilson Loops in Superconformal Chern-Simons Theories with Matter,''
  JHEP {\bf 1003}, 089 (2010)
  doi:10.1007/JHEP03(2010)089
  [arXiv:0909.4559 [hep-th]].
  %%CITATION = doi:10.1007/JHEP03(2010)089;%%


%\cite{Hanada:2012si}
\bibitem{Hanada:2012si} 
  M.~Hanada, M.~Honda, Y.~Honma, J.~Nishimura, S.~Shiba and Y.~Yoshida,
  ``Numerical studies of the ABJM theory for arbitrary N at arbitrary coupling constant,''
  JHEP {\bf 1205}, 121 (2012)
  doi:10.1007/JHEP05(2012)121
  [arXiv:1202.5300 [hep-th]].
  %%CITATION = doi:10.1007/JHEP05(2012)121;%%
  %65 citations counted in INSPIRE as of 15 Jul 2016


%\cite{Hatsuda:2015owa}
\bibitem{Hatsuda:2015owa} 
  Y.~Hatsuda and K.~Okuyama,
  ``Resummations and Non-Perturbative Corrections,''
  JHEP {\bf 1509}, 051 (2015)
  doi:10.1007/JHEP09(2015)051
  [arXiv:1505.07460 [hep-th]].
  %%CITATION = doi:10.1007/JHEP09(2015)051;%%
  %18 citations counted in INSPIRE as of 15 Jul 2016


%\cite{Assel:2015oxa}
\bibitem{Assel:2015oxa} 
  B.~Assel and J.~Gomis,
  ``Mirror Symmetry And Loop Operators,''
  JHEP {\bf 1511}, 055 (2015)
  doi:10.1007/JHEP11(2015)055
  [arXiv:1506.01718 [hep-th]].
  %%CITATION = doi:10.1007/JHEP11(2015)055;%%


%\cite{Cooke:2015ila}
\bibitem{Cooke:2015ila} 
  M.~Cooke, N.~Drukker and D.~Trancanelli,
  ``A profusion of $1/2$ BPS Wilson loops in $\mathcal{N}=4$ Chern-Simons-matter theories,''
  JHEP {\bf 1510}, 140 (2015)
  doi:10.1007/JHEP10(2015)140
  [arXiv:1506.07614 [hep-th]].
  %%CITATION = doi:10.1007/JHEP10(2015)140;%%

\bibitem{TW}
  C. A. Tracy and H. Widom, ``Proofs of Two Conjectures Related to the Thermodynamic
  Bethe Ansatz'', Commun.\ Math.\ Phys. 179 (1996) 667-680 [solv-int/9509003].


%\cite{Okuyama:2011su}
\bibitem{Okuyama:2011su} 
  K.~Okuyama,
  ``A Note on the Partition Function of ABJM theory on $S^3$,''
  Prog.\ Theor.\ Phys.\  {\bf 127}, 229 (2012)
  doi:10.1143/PTP.127.229
  [arXiv:1110.3555 [hep-th]].
  %%CITATION = doi:10.1143/PTP.127.229;%%

%\cite{Hatsuda:2012hm}
\bibitem{Hatsuda:2012hm} 
  Y.~Hatsuda, S.~Moriyama and K.~Okuyama,
  ``Exact Results on the ABJM Fermi Gas,''
  JHEP {\bf 1210}, 020 (2012)
  doi:10.1007/JHEP10(2012)020
  [arXiv:1207.4283 [hep-th]].
  %%CITATION = doi:10.1007/JHEP10(2012)020;%%


%\cite{Putrov:2012zi}
\bibitem{Putrov:2012zi} 
  P.~Putrov and M.~Yamazaki,
  ``Exact ABJM Partition Function from TBA,''
  Mod.\ Phys.\ Lett.\ A {\bf 27}, 1250200 (2012)
  doi:10.1142/S0217732312502008
  [arXiv:1207.5066 [hep-th]].
  %%CITATION = doi:10.1142/S0217732312502008;%%

%\cite{Hatsuda:2015oaa}
\bibitem{Hatsuda:2015oaa} 
  Y.~Hatsuda,
  ``Spectral zeta function and non-perturbative effects in ABJM Fermi-gas,''
  JHEP {\bf 1511}, 086 (2015)
  doi:10.1007/JHEP11(2015)086
  [arXiv:1503.07883 [hep-th]].
  %%CITATION = doi:10.1007/JHEP11(2015)086;%%

%\cite{Assel:2015hsa}
\bibitem{Assel:2015hsa} 
  B.~Assel, N.~Drukker and J.~Felix,
  ``Partition functions of 3d $\hat D$-quivers and their mirror duals from 1d free fermions,''
  JHEP {\bf 1508}, 071 (2015)
  doi:10.1007/JHEP08(2015)071
  [arXiv:1504.07636 [hep-th]].
  %%CITATION = doi:10.1007/JHEP08(2015)071;%%


%\cite{Okuyama:2015auc}
\bibitem{Okuyama:2015auc} 
  K.~Okuyama,
  ``Probing non-perturbative effects in M-theory on orientifolds,''
  JHEP {\bf 1601}, 054 (2016)
  doi:10.1007/JHEP01(2016)054
  [arXiv:1511.02635 [hep-th]].
  %%CITATION = doi:10.1007/JHEP01(2016)054;%%


%\cite{Hatsuda:2015lpa}
\bibitem{Hatsuda:2015lpa} 
  Y.~Hatsuda, M.~Honda and K.~Okuyama,
  ``Large N non-perturbative effects in $\mathcal{N}=4$ superconformal Chern-Simons theories,''
  JHEP {\bf 1509}, 046 (2015)
  doi:10.1007/JHEP09(2015)046
  [arXiv:1505.07120 [hep-th]].
  %%CITATION = doi:10.1007/JHEP09(2015)046;%%

%\cite{Okuyama:2016xke}
\bibitem{Okuyama:2016xke} 
  K.~Okuyama,
  ``Orientifolding of the ABJ Fermi gas,''
  JHEP {\bf 1603}, 008 (2016)
  doi:10.1007/JHEP03(2016)008
  [arXiv:1601.03215 [hep-th]].
  %%CITATION = doi:10.1007/JHEP03(2016)008;%%


%\cite{Okuyama:2016deu}
\bibitem{Okuyama:2016deu} 
  K.~Okuyama,
  ``Instanton Corrections of 1/6 BPS Wilson Loops in ABJM Theory,''
  JHEP {\bf 1609}, 125 (2016)
  doi:10.1007/JHEP09(2016)125
  [arXiv:1607.06157 [hep-th]].
  %%CITATION = doi:10.1007/JHEP09(2016)125;%%


\bibitem{EK1} 
  B.~Eynard and C.~Kristjansen,
  ``Exact solution of the O(n) model on a random lattice,''
  Nucl.\ Phys.\ B {\bf 455}, 577 (1995)
  doi:10.1016/0550-3213(95)00469-9
  [hep-th/9506193].
  %%CITATION = doi:10.1016/0550-3213(95)00469-9;%%

\bibitem{EK2} 
  B.~Eynard and C.~Kristjansen,
  ``More on the exact solution of the O(n) model on a random lattice and an investigation of the case $|n| > 2$,''
  Nucl.\ Phys.\ B {\bf 466}, 463 (1996)
  doi:10.1016/0550-3213(96)00104-6
  [hep-th/9512052].
  %%CITATION = doi:10.1016/0550-3213(96)00104-6;%%

%\cite{Suyama:2012uu}
\bibitem{Suyama:2012uu} 
  T.~Suyama,
  ``On Large N Solution of N=3 Chern-Simons-adjoint Theories,''
  Nucl.\ Phys.\ B {\bf 867}, 887 (2013)
  doi:10.1016/j.nuclphysb.2012.10.017
  [arXiv:1208.2096 [hep-th]].
  %%CITATION = doi:10.1016/j.nuclphysb.2012.10.017;%%


%\cite{Borot:2009ia}
\bibitem{Borot:2009ia} 
  G.~Borot and B.~Eynard,
  ``Enumeration of maps with self avoiding loops and the O(n) model on random lattices of all topologies,''
  J.\ Stat.\ Mech.\  {\bf 1101}, P01010 (2011)
  doi:10.1088/1742-5468/2011/01/P01010
  [arXiv:0910.5896 [math-ph]].
  %%CITATION = doi:10.1088/1742-5468/2011/01/P01010;%%


%\cite{Hatsuda:2013gj}
\bibitem{Hatsuda:2013gj} 
  Y.~Hatsuda, S.~Moriyama and K.~Okuyama,
  ``Instanton Bound States in ABJM Theory,''
  JHEP {\bf 1305}, 054 (2013)
  doi:10.1007/JHEP05(2013)054
  [arXiv:1301.5184 [hep-th]].
  %%CITATION = doi:10.1007/JHEP05(2013)054;%%

%\cite{Aganagic:2011mi}
\bibitem{Aganagic:2011mi} 
  M.~Aganagic, M.~C.~N.~Cheng, R.~Dijkgraaf, D.~Krefl and C.~Vafa,
  ``Quantum Geometry of Refined Topological Strings,''
  JHEP {\bf 1211}, 019 (2012)
  doi:10.1007/JHEP11(2012)019
  [arXiv:1105.0630 [hep-th]].
  %%CITATION = doi:10.1007/JHEP11(2012)019;%%

\bibitem{HMMO} 
  Y.~Hatsuda, M.~Marino, S.~Moriyama and K.~Okuyama,
  ``Non-perturbative effects and the refined topological string,''
  JHEP {\bf 1409}, 168 (2014)
  doi:10.1007/JHEP09(2014)168
  [arXiv:1306.1734 [hep-th]].
  %%CITATION = doi:10.1007/JHEP09(2014)168;%%


%\cite{Moriyama:2014gxa}
\bibitem{Moriyama:2014gxa} 
  S.~Moriyama and T.~Nosaka,
  ``Partition Functions of Superconformal Chern-Simons Theories from Fermi Gas Approach,''
  JHEP {\bf 1411}, 164 (2014)
  doi:10.1007/JHEP11(2014)164
  [arXiv:1407.4268 [hep-th]].
  %%CITATION = doi:10.1007/JHEP11(2014)164;%%


%\cite{Gu:2015pda}
\bibitem{Gu:2015pda} 
  J.~Gu, A.~Klemm, M.~Marino and J.~Reuter,
  ``Exact solutions to quantum spectral curves by topological string theory,''
  JHEP {\bf 1510}, 025 (2015)
  doi:10.1007/JHEP10(2015)025
  [arXiv:1506.09176 [hep-th]].
  %%CITATION = doi:10.1007/JHEP10(2015)025;%%

%last

 \end{thebibliography}
\end{document}